\UseRawInputEncoding
\documentclass[fleqn,pra,floatfix]{revtex4}

\usepackage{graphicx}
\usepackage{amsmath}
\usepackage{tabularx}
\usepackage{booktabs}
\usepackage{dcolumn}

\usepackage{xcolor} 

\begin{document}

\title{Hyperfine structure and electric quadrupole transitions in the deuterium molecular ion}

\author{P.~Danev}%
\email{petar_danev@abv.bg}
\affiliation{%
 Institute for Nuclear Research and Nuclear Energy, Bulgarian Academy of Sciences,\\
    blvd. Tsarigradsko ch. 72, Sofia 1142, Bulgaria}

\author{D. Bakalov}%
\affiliation{%
 Institute for Nuclear Research and Nuclear Energy, Bulgarian Academy of Sciences,\\
    blvd. Tsarigradsko ch. 72, Sofia 1142, Bulgaria}

\author{V.I.~Korobov}
\affiliation{Bogoliubov Laboratory of Theoretical Physics, Joint
Institute for Nuclear Research, 141980, Dubna, Russia}

\author{S. Schiller}%
\affiliation{%
 Institut f\"{u}r Experimentalphysik, Heinrich-Heine-Universit\"{a}t D\"{u}sseldorf, 40225 D\"{u}sseldorf, Germany}

\date{\today}

\begin{abstract}
Molecular hydrogen ions are of metrological relevance due to the
possibility of precise theoretical evaluation of their spectrum
and of external-field-induced shifts. In homonuclear molecular
ions the electric dipole $E1$ transitions are strongly suppressed,
and of primary laser spectroscopy interest is the electric
quadrupole ($E2$) transition spectrum. In continuation of previous
work on the H$_2^+$ ion, we report here the results of the
calculations of the hyperfine structure 
of the laser-induced electric quadrupole
transitions between a large set of ro-vibrational states of
D$_2^+$; the inaccuracies of previous evaluations have been corrected.
The effects of the laser polarization are studied in detail.
We show that the electric quadrupole moment of 
the deuteron can in principle be determined with low fractional uncertainty 
$(\simeq1\times10^{-4})$ by 
comparing the results presented here with future data from 
precision spectroscopy of D$_2^+$.
\end{abstract}

\maketitle


\section{Introduction}
\label{sec:H2p}

 Molecular hydrogen ions (MHIs) are three-body systems that
 offer unique possibilities for both high precision spectroscopy
 and accurate theoretical evaluation of the spectrum
 (see, e.g., Refs.~\cite{PRL2014}-\cite{science2020}
 and references therein).
 The comparison of high precision experimental
 and theoretical results opens room for independent tests of QED and 
 has the potential to provide accurate values
 of fundamental constants. Among the most impressive 
 achievements along this path is the recent determination of 
 the  proton-to-electron mass ratio with factional accuracy
 of $\sim2\times10^{-11}$ from rotational \cite{Alighanbari2018} 
 and ro-vibrational \cite{science2020} precision laser spectroscopy of 
 cooled and trapped HD$^+$ ions;
 also confirmed were the recent adjustments of the 
 CODATA values of the proton charge radius and the Rydberg constant.
 Transitions with low sensitivity to external fields are
 considered as promising candidates for the search for a time-variation of the
 mass ratios \cite{PRL2014,Karr2014}.

 In the homonuclear MHIs H$_2^+$, D$_2^+$ the electric dipole transitions are 
 strongly suppressed and the spectra are dominated by the electric quadrupole
 transitions, which are much weaker and have much smaller natural width.  
 While precision laser spectroscopy of $E2$ transitions in homonuclear 
 MHIs is still ahead in time,
 the theory has achieved significant progress.  
 The first calculations of E2 spectra 
 in the approximation of spinless particles, performed back in
 1953 by Bates and Poots \cite{BatesAndPoots}, were followed by works of
 Posen et al. \cite{Posen}, Pilon and Baye \cite{Pilon2012},
 and Pilon \cite{Pilon2013} of ever increasing precision. 
 The hyperfine structure (HFS) of the
 $E2$-spectral lines of H$_2^+$ was first considered in \cite{Karr2014};
 in Ref.~\cite{Korobov2018} the leading effects of order $O(m_e\alpha^6)$
 were also included in a systematical investigation of the spectrum.
 The HFS of the lower excited states of D$_2^+$ has been previously investigated in 
 \cite{babb-psas,Zhang2013,Zhang2016}.
 Only very few experimental studies of the HFS of D$_2^+$ have been carried out by now 
 \cite{v21,cruse}.
 
 Zhang et al. \cite{Zhang2013} and earlier Babb \cite{Bab1997,babb2} raised
 the interesting question about the possibility of determining
 $Q_d$ from the hyperfine structure of D$_2^{+}$; there is now
 an intense discussion on its actual value. Recently,
 Alighanbari et al. \cite{{Alighanbari-2020}} determined $Q_d$ with
 fractional uncertainty $u_r(Q_d)=1.5$ \% from the hyperfine spectrum of a pure
 rotational transition of HD$^+$. However, the most precise
 determination so far is from a comparison of experiment and theory
 for the neutral hydrogen molecules
 \cite{pavanello,jozwiak,komasa}. Their stated uncertainties,
 ranging from $1\times10^{-4}$ to $8\times10^{-4}$, are  so small
 that it is worthwhile to perform an independent measurement,
 using the molecular hydrogen ion HD$^+$ or D$_2^+$.

 In the present work, we apply the approach of \cite{Korobov2018} to
 the deuterium ion ${\rm D_2^+}$. In Sec. \ref{sec:hfs}, we re-evaluate
 the HFS of ${\rm D_2^+}$ in the Breit-Pauli approximation by correcting
 inaccuracies in the preceding works.
 Sec.~\ref{sec:E2} is dedicated to the study
 of the laser-stimulated $E2$ transition spectrum in ${\rm D_2^+}$
 with account of the HFS of the molecular levels and the polarization
 of the laser source. 
 In Sec.~\ref{sec:Q_d} we demonstrate that the results 
 presented here provide the theoretical input 
 for the composite frequency method  \cite{PRL2014,traceless,Alighanbari-2020} needed 
 to determine the deuteron quadrupole moment 
 from  precision spectroscopy of D$_2^+$ with 
 an uncertainty comparable with the uncertainty of the latest values 
 reported in \cite{pavanello,puchalski}.
 In the final Sec. \ref{sec:Conclusion} we summarize
 and discuss the results.

\section{Hyperfine structure of deuterium molecular ion}\label{sec:hfs}

\subsection{Theoretical model}\label{subs2.1}

 The nonrelativistic Hamiltonian of the hydrogen molecular ion ${\rm D}_2^+$ is:
 \begin{equation}\label{nonrelhamilt}
 H^{\rm NR} =
   \frac{\mathbf{p}_1^2}{2m_d}+\frac{\mathbf{p}_2^2}{2m_d}+\frac{\mathbf{p}_e^2}{2m_e}
   +\frac{e^2}{4\pi \varepsilon_0}\left(-\frac{1}{r_1}-\frac{1}{r_2}+\frac{1}{r_{12}}\right),
\end{equation}
 where $m_{d}$ and $m_e$ are the masses of the deuterons and the electron,
 $\mathbf{R}_1$, $\mathbf{R}_2$, $\mathbf{R}_e$ and $\mathbf{p}_1$,
 $\mathbf{p}_2$, $\mathbf{p}_e$ are the position and momentum vectors of
 the two deuterons and the electron in the center of mass frame,
 and $\mathbf{r}_{1,2}=\mathbf{R}_e\!-\!\mathbf{R}_{1,2}$,
 $\mathbf{r}_{12}=\mathbf{R}_2\!-\!\mathbf{R}_1$,
 $r_{1,2}=|{\bf r}_{1,2}|$,  $r_{12}=|{\bf r}_{12}|$.
 We consider only $\Sigma_g$ states of D$^+_2$; in the nonrelativistic
 approximation the discrete $\Sigma_g$ states of the hydrogen isotope
 molecular ions are labeled with the quantum numbers of the nuclear
 vibrational excitation $v$, of the total orbital momentum $L$,
 and of its projection on the space-fixed quantization axis $L_z$;
 the spatial parity $\lambda=\pm1$ is constrained to $\lambda=(-1)^L$
 and will be omitted in further notations. The nonrelativistic
 (Coulomb) energy levels and wave functions of D$_2^+$
 in the state $|vLL_z\rangle$ are denoted by
 $E^{{\rm (NR)}vL}$ and $\Psi^{{\rm (NR)}vLL_z}$, respectively.

  The leading-order spin effects
  are described by adding to $H^{\rm NR}$
  the pairwise spin interaction terms $V$ of the Breit-Pauli Hamiltonian
  of Ref.~\cite{Bakalov2006}:
\begin{equation}
H=H^{\rm NR}+V,
\qquad
V=V_{ed_1}+V_{ed_2}+V_{dd},
  \label{fullH}
\end{equation}
where $d_1$ and $d_2$ denote the two deuterium nuclei of D$_2^+$.
We remind the explicit form of the spin interaction operators; to
comply with the established traditions we shall use atomic units
$\hbar=e=4\pi\varepsilon_0=1$ in the remainder of
Sect.~\ref{subs2.1}.
\begin{equation}\label{eq:Ved}
\begin{array}{@{}l}\displaystyle
V_{ed_1} = \alpha^2
   \biggl[
      -\frac{4 \pi}{3}\,\mu_e \mu_d\,\frac{m_e}{m_p}\,\delta({\bf r}_1)\,({\bf s}_e \cdot {\bf I}_1)
      +\left(\mu_e-\frac{1}{2}\right)\frac{1}{r_1^3}({\bf r}_1\times {\bf p}_e)\cdot{\bf s}_e
      -\mu_e\frac{m_e}{m_d}\frac{1}{r_1^3}({\bf r}_1\times {\bf p}_1)\cdot{\bf s}_e
\\[3mm]\displaystyle\hspace{12mm}
      -\frac{m_e}{2m_d}\left(\mu_d\frac{m_e}{m_p}-\frac{m_e}{m_d}\right)
                                \frac{1}{r_1^3}({\bf r}_1\times {\bf p}_1)\cdot{\bf I}_1
      +\mu_d\frac{m_e}{2m_p}\frac{1}{r_1^3}({\bf r}_1\times {\bf p}_e)\cdot{\bf I}_1
  \biggr]
\\[3mm]\displaystyle\hspace{12mm}
   +\alpha^2
   \left[
      \mu_e \mu_d\,\frac{m_e}{2 m_p}\frac{r_1^2({\bf s}_e \cdot {\bf I}_1)
      -3({\bf r}_1 \cdot {\bf s}_e)({\bf r}_1 \cdot {\bf I}_1)}{r_1^5}
   \right]
   +\frac{Q_d}{2a_0^2}\,\frac{r_1^2{\bf I}_1^2-3({\bf r}_1\cdot {\bf I}_1)^2}{r_1^5}\,,
\end{array}
\end{equation}
\begin{equation}
V_{ed_2} = V_{ed_1}(\mathbf{r}_1\to\mathbf{r}_2,
                    \mathbf{p}_1\to\mathbf{p}_2,
                    \mathbf{I}_1\to\mathbf{I}_2),
\end{equation}
\begin{equation}\label{eq:Vdd}
\begin{array}{@{}l}\displaystyle
V_{dd} = \alpha^2
   \biggl[
      -\frac{2\pi}{3}\mu_d^2\,\frac{m_e^2}{m_p^2}\,\delta({\bf r}_{12})\,({\bf I}_1 \cdot {\bf I}_2)
      +\frac{\mu_d m_e^2}{2m_p m_d}\left(\frac{1}{r_{12}^3}({\bf r}_{12}\times {\bf p}_1)\cdot{\bf I}_2
      -\frac{1}{r_{12}^3}({\bf r}_{12}\times{\bf p}_2)\cdot{\bf I}_1\right)
\\[3mm]\displaystyle\hspace{12mm}
      +\frac{m_e}{2m_d}\left(\mu_d\frac{m_e}{m_p}-\frac{m_e}{m_d}\right)
      \left(
         \frac{1}{r_{12}^3}({\bf r}_{12}\times {\bf p}_1)\cdot{\bf I}_1
         -\frac{1}{r_{12}^3}({\bf r}_{12}\times {\bf p}_2)\cdot{\bf I}_2
      \right)
   \biggr]
\\[3.5mm]\displaystyle\hspace{12mm}
   +\alpha^2
      \left[
         \mu_d^2\,\frac{m_e^2}{4m_p^2}
         \frac{r_{12}^2({\bf I}_1 \cdot {\bf I}_2)-3({\bf r}_{12} \cdot {\bf I}_1)({\bf r}_{12} \cdot {\bf I}_2)}
                                                                                                       {r_{12}^5}
      \right]
   -\frac{Q_d}{2a_0^2}\sum\limits_{i=1,2}\frac{r_{12}^2{\bf I}_i^2-3({\bf r}_{12}\cdot{\bf I}_i)^2}{r_{12}^5}\,.
\end{array}
\end{equation}
 Here $\mathbf{I}_{1,2}$ and $\mathbf s_e$ are the
 spin operators of the two deuterons and of the electron,
 respectively, and proper symmetrization of the terms containing
 non-commuting operators is assumed;
 $\mu_e $ is the magnetic dipole moment of the electron in units
 $\mu_0=e\hbar/2m_e$ (Bohr magneton), $\mu_d$ is the magnetic dipole
 moment of the deuteron in units $\mu_N=e\hbar/2m_p$ (nuclear magneton),
 $a_0$ is the Bohr radius,
 and $Q_d$ is the electric quadrupole moment of the deuteron.
 The same spin-interaction Hamiltonian was used in
 \cite{Zhang2013}.

 The spin interactions split the degenerate nonrelativistic
 energy levels $E^{{\rm (NR)}vL}$ into a manifold of hyperfine
 levels that are distinguished with additional quantum numbers
 (QNs) describing their ``spin composition''.
As evidenced in subsection \ref{sec:hfsnum}, the appropriate angular momentum coupling scheme for D$_2^+$ is
\[
\mathbf{I} = \mathbf{I}_1+\mathbf{I}_2,
\qquad
\mathbf{F}=\mathbf{I}+\mathbf{s}_e,
\qquad
\mathbf{J}=\mathbf{L}+\mathbf{F}.
\]
Accordingly, the hyperfine states are labelled with the exact QNs of the total angular momentum $J$ and its projection $J_z$, and the approximate QNs $I$, $F$ and $L$.
 Similar to Refs.~\cite{Bakalov2006,Zhang2013,Korobov2018},
 in first order of perturbation theory the hyperfine levels
 $E^{(vL)IFJJ_z}$ may be put in the form
 $E^{(vL)IFJJ_z}=E^{{\rm (NR)}vL}+\Delta E^{(vL)IFJJ_z}$,
 where the corrections $\Delta E^{(vL)IFJJ_z}$, also referred to as
  'hyperfine energies' or 'hyperfine shifts',
 are the eigenvalues of the effective
 spin interaction Hamiltonian $H^{\rm eff}$.
 (Of course, in absence of external fields the energies are
 degenerate in $J_z$).
 Instead of the above, however, we shall use the more general form
 \begin{equation}
 E^{(vL)IFJJ_z}=E^{{\rm (diag)}vL}+\Delta E^{(vL)IFJJ_z},
 \label{eq:hfsspl}
 \end{equation}
 where $E^{{\rm (diag)}vL}$ also includes the relativistic, QED, etc. 
 spin-independent corrections to the non-relativistic energy levels $E^{{\rm (NR)}vL}$. 
 Since the evaluation of $\Delta E^{(vL)IFJ}$ is the point where our results disagree to some extent with the
 results of Refs.~\cite{Zhang2013, Zhang2016}, we give more details of the
 calculations.

 \subsection{Effective spin Hamiltonian}

 \begin{figure}[t]
 \begin{center}
 \includegraphics[width=0.9\textwidth]{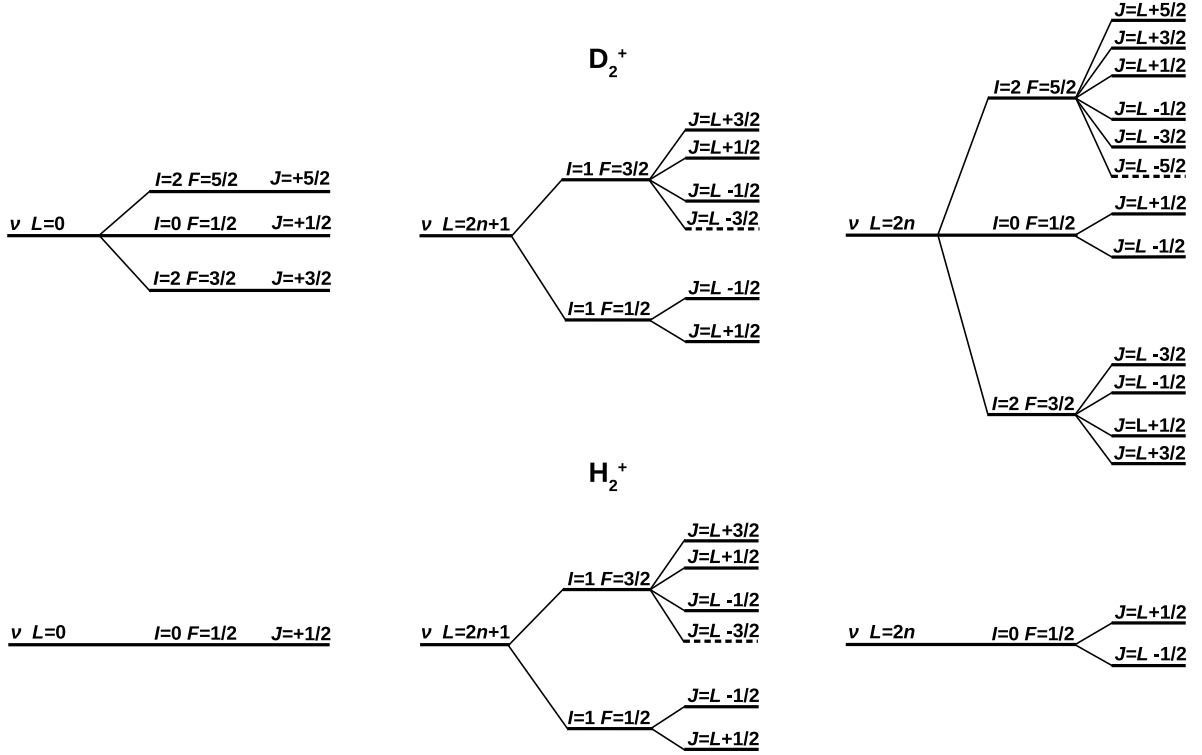}
 \caption{Comparison of the hyperfine structure of the molecular
 ions D$_2^+$ and H$_2^+$, for even $L=2n, n=1,2,\ldots$ and odd
 $L=2n+1, n=0,1,\ldots$ values of $L$ and for $L\!=\!0$. For $1\le
 L\le2$ the states marked with dashed lines do not exist (see
 Tables \ref{tab:odd},\ref{tab:even}).
 For $L\!=\!0$ the HFS of D$_2^+$ consists of only
 three states; similar to H$_2^+$ the one with $I\!=\!0$ has no hyperfine shift.}
 \label{Fig:D2p_hfs}
 \end{center}
 \end{figure}

We associate with the spin of the deuteron $d_1$ the $(2I_1+1)$-dimensional space of the irreducible representation $(I_1)$ of $su(2)$ with basis vectors $|I_1I_{1z}\rangle,I_{1z}=-I_1,\ldots,I_1$, satisfying
\[
\hat{\mathbf{I}}_1^2|I_1I_{1z}\rangle=I_1(I_1\!+\!1)|I_1I_{1z}\rangle,
\qquad
\hat{I}_{1z}|I_1I_{1z}\rangle=I_{1z}|I_1I_{1z}\rangle.
\]
 Similarly, we define the sets
 $|I_2I_{2z}\rangle$, $|s_es_{ez}\rangle$,
 and $|LL_z\rangle$. The basis set $|IFF_z\rangle$ in the
 resulting space of the spin variables of all D$_2^+$
 constituents is taken in the form:
 \begin{equation}
 |IFF_z\rangle=\sum\limits_{I_{1z} I_{2z}s_{ez}I_z}
 C_{I_1I_{1z},I_2I_{2z}}^{II_z}
 C_{II_z,s_es_{ez}}^{FF_z}
 |I_1I_{1z}\rangle|I_2I_{2z}\rangle|s_es_{ez}\rangle,
 \label{eq:su1}
 \end{equation}
 here $C_{l_1m_1,l_2m_2}^{LM}=\left\langle l_1l_2m_1m_2|LM\right\rangle$ are Clebsch-Gordan coefficients.
 Thus, the basis set in the hyperfine manifold of the $(vL)$ state of D$_2^+$ consists of the functions
 \begin{equation}
 \Psi_0^{(vL)IFJJ_z}=\sum\limits_{L_zF_z}C_{LL_z,FF_z}^{JJ_z}\,
 \Psi^{{\rm (NR)}vLL_z}\,|IFF_z\rangle.
 \label{eq:degenPT}
 \end{equation}
 We also define a basis set depending on the angular part only, which will be referred to as "pure" states,
 \begin{equation}\label{eq:pure_states}
 |LIFJJ_z\rangle=\sum\limits_{L_z,F_z}
 C_{LL_z,FF_z}^{JJ_z}
 |LL_z\rangle|IFF_z\rangle.
 \end{equation}
 The effective spin Hamiltonian $H^{\rm eff}$ is a
 matrix operator acting on the finite-dimensional space spanned by
 the vectors $|LIFJJ_z\rangle$, such that
\begin{equation}\label{eq:vmatdef}
H^{{\rm eff}(vL)J}_{I'F',IF} \equiv
   \langle LI'F'JJ_z|H^{\rm eff}|LIFJJ_z\rangle =
   \langle \Psi_0^{(vL)I'F'JJ_z}|V|\Psi_0^{(vL)IFJJ_z}\rangle.
\end{equation}
In absence of external fields the matrix of $H^{{\rm eff}(vL)J}_{I'F',IF}$ is independent of $J_z$. For the deuterium molecular ion $H^{\rm eff}$ has the form
\begin{equation}\label{eq:heff}
\begin{array}{@{}l}\displaystyle
H^{\rm eff} =
   E_1(\mathbf{L}\cdot \mathbf{s}_e)
   +E_2(\mathbf{L}\cdot \mathbf{I})
   +E_3(\mathbf{I}\cdot \mathbf{s}_e)
\\[2mm]\displaystyle\hspace{11mm}
   +E_4\bigl(2\mathbf{L}^2(\mathbf{I}\cdot \mathbf{s}_e)
   -3\left(
   (\mathbf{L}\cdot \mathbf{I})(\mathbf{L}\cdot \mathbf{s}_e)
   +(\mathbf{L}\cdot \mathbf{s}_e)(\mathbf{L}\cdot \mathbf{I})
   \right)
   \bigr)
\\[2.5mm]\displaystyle\hspace{11mm}
   +E_5
   \big(
   2\mathbf{L}^2(\mathbf{I}_1\cdot \mathbf{I}_2)-3
   \left((\mathbf{L}\cdot \mathbf{I}_1)(\mathbf{L}\cdot \mathbf{I}_2)
   +(\mathbf{L}\cdot \mathbf{I}_2)(\mathbf{L}\cdot \mathbf{I}_1)\right)
   \big)
\\[0.5mm]\displaystyle\hspace{11mm}
   +E_6\left[{\bf L}^2{\bf I}_1^2-\frac{3}{2}({\bf L}\cdot{\bf I}_1)-3({\bf L}\cdot{\bf I}_1)^2
   +{\bf L}^2{\bf I}_2^2-\frac{3}{2}({\bf L}\cdot{\bf I}_2)-3({\bf L}\cdot{\bf I}_2)^2
   \right].
\end{array}
\end{equation}
 Compared to the effective spin Hamiltonian for H$^+_2$ of
 Ref.~\cite{Korobov2018}, $H^{\rm eff}$ of Eq.~(\ref{eq:heff})
 includes one additional term (with $E_6$, last line) that
 describes the effects due to the electric quadrupole moment of
 the nuclei and arises when averaging the last term in
 Eqs.~(\ref{eq:Ved}-\ref{eq:Vdd}).
 The first four terms in Eq.~(\ref{eq:heff}) coincide
 with the first four terms in the effective Hamiltonian of
 Refs.~\cite{Zhang2013,Zhang2016}, defined in Eqs.~(12)-(15) of the former,
 with account of the correspondence between the notations used:
 $E_1=c_e$, $E_2=c_I$, $E_3=b_F$, $E_4=d_1/(3(2L-1)(2L+3))$.
 The last two terms involving $E_5$ and $E_6$, however, do not. The
 disagreement appears in the terms related to the
 tensor interaction of the deuterons in
 the last lines of Eqs.~(\ref{eq:Ved}) and (\ref{eq:Vdd}). The explicit
 expressions for $E_4$--$E_6$ are:
\begin{equation}
\begin{array}{@{}l}\displaystyle
E_4 =
   \alpha^2\mu_e\mu_d\frac{m_e}{2m_p}
   \left[
      \frac{\langle vL\|r_1^{-5}(\mathbf{r}_1^2\delta_{i\!j}-3r_{1i}r_{1j})\|vL\rangle}
           {\langle L\|2\mathbf{L}^2\delta_{i\!j}-3(L_iL_j\!+\!L_jL_i)\|L\rangle}
      +(1\to2)
   \right],
\\[3mm]\displaystyle
E_5 =
   \alpha^2\left(\frac{\mu_dm_e}{2m_p}\right)^2
   \frac{\langle vL\|r_{12}^{-5}(\delta_{i\!j}\mathbf{r}_{12}^2-3r_{12i}r_{12j})\|vL\rangle}
      {\langle L\|2\mathbf{L}^2\delta_{i\!j}-3(L_iL_j\!+\!L_jL_i)\|L\rangle}\,,
\\[3mm]\displaystyle
E_6 =Q_d \bar{E}_6,\ \bar{E}_6=
   \frac{1}{2}
   \left[
      \frac{\langle vL\|r_1^{-5}(\delta_{i\!j}\mathbf{r}_1^2-3r_ir_j)\|vL\rangle
           -\langle vL\|r_{12}^{-5}(\delta_{i\!j}\mathbf{r}_{12}^2-3r_{12i}r_{12j})\|vL\rangle}
                                       {\langle L\|\mathbf{L}^2\delta_{i\!j}-(3/2)(L_iL_j\!+\!L_jL_i)\|L\rangle}
      +(1\to2)
   \right],
\end{array}
\label{eq:E456}
\end{equation}
 where $\langle v'L'\|\ldots\|vL\rangle$ denote reduced matrix
 elements in the basis of nonrelativistic wave functions
 $\psi^{{\rm (NR)}vLL_z}$, while
 \[
 \langle L\|2\mathbf{L}^2\delta_{ij}-3(L_iL_j\!+\!L_jL_i)\|L\rangle =
 -2\sqrt{L(L\!+\!1)}\sqrt{(2L\!-\!1)(2L\!+\!1)(2L\!+\!3)}.
 \]
 is a reduced matrix element in the basis of the representation
 $(L)$ of $su(2)$ (see Eqs.~(\ref{eq:su1}),(\ref{eq:pure_states})).
 The point is that the tensor terms in $H^{\rm eff}$ have non-zero matrix elements
 between ''pure'' states states with even $L$ and different values of the total nuclear spin $I=0$
 and $I'=2$: 
 \begin{align}
 &\langle \nu LIFJ\|H^{\rm eff} \|\nu LI'F'J\rangle = 
 10\sqrt{3}(-1)^{J+2F+L+1/2}\\
 &\times \sqrt{(2F+1)(2F'+1)} L(L+1)(2L+1)\left\{\begin{matrix}
      F' & \!F\! & 2 \\
      L & \!L\! & J
   \end{matrix}\right\}
    \left\{\begin{matrix}
     1 & \!1\! & 2 \\
      L & \!L\! & L
   \end{matrix}\right\}
       \left\{\begin{matrix}
     I & \!1/2\! & F \\
      F' & \!2\! & I'
   \end{matrix}\right\}\left( E_5-E_6
   \right),\nonumber
  \end{align}
 while 
 the effective Hamiltonian 
 of Refs.~\cite{Zhang2013, Zhang2016} is {\em diagonal} in
 $\mathbf{I}$. Neglecting the coupling of 
 pure states with different $I$ affects the values of the hyperfine shifts at 
 the kHz level, as illustrated in Table~\ref{tab:comparison} of Section~\ref{sec:hfsnum}.

 In first order of perturbation theory the wave functions of
 D$_2^+$ are linear
 combinations of the basis set (\ref{eq:degenPT}):
 \begin{equation}
 \Psi^{(vL)IFJJ_z}=\sum\limits_{I'F'}\beta^{(vL)IFJ}_{I'F'}\Psi^{(vL)I'F'JJ_z}_{0},
 \label{eq:betaexpans}
 \end{equation}
 where the constant amplitudes $\beta^{(vL)IFJ}_{I'F'}$ and the hyperfine
 shifts $\Delta E^{(vL)IFJ}$ are the eigenvectors
 and eigenvalues of the matrix of $H^{\rm eff}$
 in the basis of pure states (\ref{eq:pure_states}):
 \begin{equation}
 \sum\limits_{I''F''}H^{{\rm eff}(vL)J}_{I'F',I''F''}\
 \beta^{(vL)IFJ}_{I''F''}=
 \Delta E^{(vL)IFJ}\beta^{(vL)IFJ}_{I'F'}.
 \label{eq:eigprob}
 \end{equation}
 Similar to H$_2^+$, symmetry with respect to exchange of
 the identical nuclei imposes restrictions
 on the allowed values of $I$ in the summation in Eqs.~(\ref{eq:betaexpans}),
 (\ref{eq:eigprob}); for $\Sigma_g$ states, in
 particular, $I$ must satisfy $(-1)^{L+I}=1$. As a result,
 in H$^+_2$, in first order of perturbation theory, $I$ turns out
 to be an exact quantum number \cite{Korobov2018}.
 This is not the case for D$_2^+$, however, where both values
 $I=0$ and $I=2$ are allowed for even values
 of $L$, although their mixing is weak.
 Still, a few of the hyperfine states  of D$_2^+$ are ``pure
 states'' with no mixing and all quantum numbers {\em exact};
 these are the states with $I=1$, $F=3/2$, $J=L\pm F$
 (for odd $L$), the states with $I=2$, $F=5/2$, $J=L\pm F$ (for even $L\ge2$),
 as well as the states with $L=0$ and either
 $I=0, J=F=1/2$ or $I=2, J=F=I\pm1/2$.
 The ``stretched'' states with $F=I+1/2$, $J=L+F$, and
 $J_z=\pm J$, of significant experimental interest
 because their Zeeman shift is strictly linear in the magnetic field
 in first order of perturbation theory \cite{jpb44}, are a sub-class of
 the ``pure'' states listed above.
 The hyperfine energy $\Delta E\equiv\Delta E^{(vL)IFJ}$
 of the ``pure'' states is simply expressed in terms of
 the coefficients of the effective spin Hamiltonian as follows:
 \begin{eqnarray*}
 &&\Delta E^{(vL)IFJ}=\frac{1}{2}\big( E_3+L(E_1+2E_2+
 (2L-1)(E_6-2E_4-2E_5))\big),\ \text{for}\ I=1,F=3/2,J=L+F,\
 \text{odd}\ L;\\
 &&\Delta E^{(vL)IFJ}=\frac{1}{2}\big( E_3-(L+1)(E_1+2E_2+
 (2L+3)(E_6-2E_4-2E_5))\big),\ \text{for}\ I=1,F=3/2,J=L-F,\
 \text{odd}\ L\ge3;\\
 &&\Delta E^{(vL)IFJ}= E_3+L(E_1/2+2E_2-
 (2L-1)(2E_4+2E_5+E_6)),\ \text{for}\ I=2,F=5/2,J=L+F,\
 \text{even}\ L\ge2;\\
 &&\Delta E^{(vL)IFJ}= E_3-(L+1)(E_1/2+2E_2+
 (2L+3)(2E_4+2E_5+E_6)),\ \text{for}\ I=2,F=5/2,J=L-F,\
 \text{even}\ L\ge4;\\
 &&\Delta E^{(v0)0FJ}=0\ \text{for}\ J=F=1/2,\\
 &&\Delta E^{(v0)2FJ}=-3E_3/2\ \text{for}\ J=F=3/2,\\
 &&\Delta E^{(v0)2FJ}=E_3\ \text{for}\ J=F=5/2.
 \end{eqnarray*}

 \subsection{Numerical results}
 \label{sec:hfsnum}

The numerical results of the present work were obtained using the
non-relativistic wave functions $\Psi^{{\rm (NR)}vLL_z}$ of
D$^+_2$, calculated with high numerical precision in the
variational approach of Ref.~\cite{KorobovVarMethod}. Throughout
the calculations the CODATA18 values \cite{nist} of fundamental
constants were used for $m_d/m_p$, $\mu_e$, and $\mu_d$, while
$Q_d=0.285783\text{ fm}^2$ was taken from \cite{pavanello}.
 Table~\ref{tab:effs} gives the values of $E_n,n=1,\ldots,6$
 for the lower ro-vibrational states of D$_2^+$; the values
 for all ro-vibrational states $(vL)$ with $v\le10$ and
 $L\le4$ can be found in the electronic supplement \cite{suppl}.
 We list the numerical values of $E_n,n=1,\ldots,6$ with 6 significant
 digits to avoid rounding errors in further calculations, 
 but one should keep in mind that  
 the contribution from QED and relativistic effects of
 order $O(m_e\alpha^6)$ and higher is not accounted for by
 the Breit-Pauli Hamiltonian. 
 We denote by $u(E_n)$ and $u_r(E_n)$ the absolute and 
 fractional uncertainties of $E_n$;
 $u_r(E_n),n\le5$ are therefore estimated to be order
 $O(\alpha^2)\sim0.5\times10^{-4}$ (for $u_r(E_6)$ see Eq.~(\ref{eq:uncE6}) below). The numerical uncertainty stemming from
 numerical integration of the non-relativistic wave functions 
 $\Psi^{{\rm (NR)}vLL_z}$ is smaller. Also smaller is the contribution
 to $u_r(E_n)$ from the uncertainty in the values of the physical constants 
 in the Breit-Pauli Hamiltonian (\ref{eq:Ved}-\ref{eq:Vdd}), with the important exception of 
 $E_6$, for which the uncertainty $u(Q_d)$ of the electric quadrupole moment 
 contributes substantially. 
 Indeed, from Eq.~(\ref{eq:E456}) we obtain 
 \begin{equation}
 u_r(E_6)=\frac{u(E_6)}{E_6}=\sqrt{u_r^2(Q_d)+u_r^2(\bar{E}_6)}\text{\ with\ }
 u_r(\bar{E}_6)\sim u_r(E_n)\sim O(\alpha^2),n=1,...,5.
 \label{eq:uncE6}
 \end{equation} 
 The uncertainty of $Q_d$ arises from uncertainties of both the experimental data 
 and the underlying molecular theory (when $Q_d$ is determined by molecular spectroscopy). 
 The relation between $u_r(Q_d)$ and the theoretical uncertainties 
 in the case of D$_2^+$ spectroscopy is treated in detail in Sec.~\ref{sec:Q_d}.

\begin{table}
\begin{tabular}{c@{\hspace{2mm}}c@{\hspace{2mm}}d@{\hspace{-4mm}}d@{\hspace{7mm}}
                d@{\hspace{-4mm}}d@{\hspace{-4mm}}d@{\hspace{-4mm}}d}
\hline\hline
\vrule width 0pt height 9.5pt 
    $v$ & $L$
  & \multicolumn{1}{r}{$E_1/h$\hspace*{-1mm}}
  & \multicolumn{1}{r}{$E_2/h$\hspace*{14mm}}
  & \multicolumn{1}{r}{$E_3/h$\hspace*{-1mm}}
  & \multicolumn{1}{r}{$E_4/h$\hspace*{4mm}}
  & \multicolumn{1}{r}{$E_5/h$\hspace*{4mm}}
  & \multicolumn{1}{r}{$E_6/h$\hspace*{7mm}}\\
\hline
\vrule width 0pt height 10pt 
 0 &  0 &    &                           &    142.533 \\
 1 &  0 &    &                           &    139.837 \\
 2 &  0 &    &                           &    137.286 \\
 3 &  0 &    &                           &    134.873 \\
 4 &  0 &    &                           &    132.593 \\
 0 &  1 &    21.4599 &   -3.21518[-3] &    142.448 &    1.32980    &   -4.71443[-4] &    5.67068[-3] \\
 1 &  1 &    20.5231 &   -3.12947[-3] &    139.756 &    1.26978    &   -4.57403[-4] &    5.66924[-3] \\
 2 &  1 &    19.6183 &   -3.04185[-3] &    137.209 &    1.21191    &   -4.43279[-4] &    5.64719[-3] \\
 3 &  1 &    18.7423 &   -2.95244[-3] &    134.799 &    1.15600    &   -4.29061[-4] &    5.60575[-3] \\
 4 &  1 &    17.8923 &   -2.86128[-3] &    132.522 &    1.10187    &   -4.14738[-4] &    5.54600[-3] \\
 0 &  2 &    21.3955 &   -3.20207[-3] &    142.278 &    3.15886[-1] &   -1.11799[-4] &    1.34038[-3] \\
 1 &  2 &    20.4612 &   -3.11645[-3] &    139.594 &    3.01623[-1] &   -1.08462[-4] &    1.34003[-3] \\
 2 &  2 &    19.5586 &   -3.02901[-3] &    137.054 &    2.87871[-1] &   -1.05106[-4] &    1.33480[-3] \\
 3 &  2 &    18.6847 &   -2.93977[-3] &    134.652 &    2.74583[-1] &   -1.01728[-4] &    1.32498[-3] \\
 4 &  2 &    17.8368 &   -2.84875[-3] &    132.382 &    2.61720[-1] &   -9.83254[-5] &    1.31082[-3] \\
 0 &  3 &    21.2995 &   -3.18256[-3] &    142.025 &    1.46902[-1] &   -5.18607[-5] &    6.18729[-4] \\
 1 &  3 &    20.3688 &   -3.09714[-3] &    139.353 &    1.40266[-1] &   -5.03083[-5] &    6.18561[-4] \\
 2 &  3 &    19.4696 &   -3.00991[-3] &    136.824 &    1.33867[-1] &   -4.87468[-5] &    6.16135[-4] \\
 3 &  3 &    18.5989 &   -2.92092[-3] &    134.432 &    1.27684[-1] &   -4.71753[-5] &    6.11585[-4] \\
 4 &  3 &    17.7541 &   -2.83015[-3] &    132.173 &    1.21697[-1] &   -4.55926[-5] &    6.05027[-4] \\
 0 &  4 &    21.1727 &   -3.15685[-3] &    141.691 &    8.54566[-2] &   -3.00678[-5] &    3.56385[-4] \\
 1 &  4 &    20.2467 &   -3.07169[-3] &    139.033 &    8.15938[-2] &   -2.91641[-5] &    3.56285[-4] \\
 2 &  4 &    19.3519 &   -2.98476[-3] &    136.519 &    7.78684[-2] &   -2.82552[-5] &    3.54880[-4] \\
 3 &  4 &    18.4855 &   -2.89608[-3] &    134.142 &    7.42682[-2] &   -2.73407[-5] &    3.52246[-4] \\
 4 &  4 &    17.6446 &   -2.80567[-3] &    131.896 &    7.07820[-2] &   -2.64197[-5] &    3.48452[-4] \\
\hline\hline
\end{tabular}
\caption{Coefficients of the effective spin interaction
Hamiltonian of Eq.~(\ref{eq:heff}) for the lower ro-vibrational
states of D$^+_2$ with $L\le4$, $v\le10$, in MHz. The number in
brackets denote powers of ten: $a[b]=a\times10^b$. Note that for
$L=0$ all coefficients but $E_3$ are zero.} \label{tab:effs}
\end{table}

 The hyperfine energies $\Delta E^{(vL)IFJ}$ and the amplitudes $\beta^{(vL)IFJ}_{I'F'}$
 of the states from the hyperfine structure of the lower excited states, 
 calculated using Eqs.~(\ref{eq:vmatdef}), (\ref{eq:heff}),
 and (\ref{eq:eigprob}), are given in
 Tables~\ref{tab:odd} and \ref{tab:even}. The results for the
 higher excited states with $v$ up to 10 or $L$ up to 4 are
 available in the electronic supplement \cite{suppl}.
 The theoretical uncertainties $u(\Delta E^{(vL)IFJ})$ of the hyperfine energies
 $\Delta E^{(vL)IFJ}$ were estimated in the assumption that the uncertainties of 
 the coefficients $E_n$ are uncorrelated, and are given by
 \begin{equation}
 u(\Delta E^{(vL)IFJ})=
 \sqrt{\sum_n \left(
 u_r(E_n)\
 \Gamma^{(vL)IFJ}_n
 \right)^2}\text{\ with\ }
 \Gamma^{(vL)IFJ}_n=E_n\frac{\partial \Delta E^{(vL)IFJ}}{\partial E_n}.
 \label{eq:uncDEhfs}
 \end{equation}
  For states with the
 smallest shifts ($\simeq20\,$ MHz, in which the influence of
 the coefficient $E_3$ is comparatively small),  the
 uncertainties $u(\Delta E^{(vL)IFJ})$ are of order 1\,kHz, while for hyperfine states with shifts 
 of the order of 100\,MHz and above the theoretical
 uncertainty may reach $\sim10$\,kHz.

 In the lower ro-vibrational states of D$_2^+$ the dominating term of the
 Breit-Pauli Hamiltonian (\ref{eq:Ved}-\ref{eq:Vdd}) is the contact spin-spin
 interaction between the electron and the nuclei; this can be
 recognized by comparing the value of $E_3$ with the other
 coefficients of the effective spin Hamiltonian $H^{\rm eff}$
 (\ref{eq:heff}), given in Table~\ref{tab:effs} and the electronic
 supplement  \cite{suppl}.
 The contribution to $\Delta E^{(vL)IFJ}$ of the $E_3$-term alone is
 $(E_3/2)(F(F+1)-I(I+1)-3/4)$;
 similar to H$_2^+$ it qualitatively determines the shape of the
 hyperfine level structure (see Fig.~\ref{Fig:D2p_hfs}), and for
 $L=0$ this is the only contribution to the hyperfine energy.
 The typical separation between
 hyperfine levels of  D$_2^+$ with different values of $F$ or $I$
 is of the order of
 $E_3\sim 10^2$ MHz.
 It is significantly smaller than the GHz separation in H$_2^+$
 or HD$^+$ because of the smaller magnetic dipole moment of the
 deuteron $\mu_d$ as compared with $\mu_p$.
 For all three molecular ions the separation between states with
 $\Delta J=\pm1$ is of the order of 10 MHz.
 The  off-diagonal elements of the
 matrix $\beta^{IFJ}_{I'F'}$ (see  Eq.~(\ref{eq:betaexpans}))
 are small, i.e. the mixing of states with different
 values of $I$ or $F$ is weak. This justifies our choice of the
 angular momentum coupling scheme, and allows to use in
 estimates of the characteristics of D$_2^+$ (except for the
 hyperfine shifts) the approximation of
 pure states $\beta^{IFJ}_{I'F'}=\delta_{F'F}\delta_{I'I}$.

 \begin{table}
 \begin{tabular}{@{\hspace{2mm}}c@{\hspace{3mm}}c@{\hspace{3mm}}c@{\hspace{5mm}}
 r@{\hspace{5mm}}r@{\hspace{5mm}}r@{\hspace{5mm}}r@{\hspace{3mm}}}
 \hline\hline
\vrule width 0pt height 10pt depth 2.5pt
 $I$ & $F$ & $J$ &  \multicolumn{1}{c}{$\Delta E^{\rm hfs}/h$} &
 \multicolumn{1}{c}{$\beta^{IFJ}_{1,1/2}$} &
 \multicolumn{1}{c}{$\beta^{IFJ}_{1,3/2}$} &
 \multicolumn{1}{c}{$\left(d\Delta E^{\rm hfs}/dQ_d\right)/h$} \\
\hline
 \multicolumn{7}{c}{$v=0$, $L=1$} \\
\hline
  1 & 1/2 & 3/2 & -146.999(8) &    0.99776 &   -0.06688 &    2.78 \\
  1 & 1/2 & 1/2 & -136.493(7) &    0.99673 &   -0.08075 &   -10.97 \\
  1 & 3/2 & 1/2 &   47.916(4) &    0.08075 &    0.99673 &    60.58 \\
  1 & 3/2 & 3/2 &   \textrm{70.351(4)} &    \textrm{0.06688} &    \textrm{0.99776} &   \textrm{-42.47} \\
  {\bf 1} & {\bf 3/2} & {\bf 5/2} & \textbf{80.624(4)} &
  \textbf{0.00000} & \textbf{1.00000} & \textbf{9.92} \\
\hline
 \multicolumn{7}{c}{$v=1$, $L=1$} \\
\hline
  1 & 1/2 & 3/2 & -144.085(7) &    0.99787 &   -0.06524 &    2.72 \\
  1 & 1/2 & 1/2 & -134.026(7) &    0.99692 &   -0.07838 &   -10.66 \\
  1 & 3/2 & 1/2 &   47.564(4) &    0.07838 &    0.99692 &    60.25 \\
  1 & 3/2 & 3/2 &   69.012(4) &    0.06524 &    0.99787 &   -42.39 \\
  {\bf 1} & {\bf 3/2} & {\bf 5/2} &   \textbf{78.870(4)} &    \textbf{0.00000} &
  \textbf{1.00000} &    \textbf{9.92} \\
\hline
 \multicolumn{7}{c}{$v=2$, $L=1$} \\
  1 & 1/2 & 3/2 & -141.325(7) &    0.99798 &   -0.06356 &  2.64 \\
  1 & 1/2 & 1/2 & -131.698(7) &    0.99711 &   -0.07598 & -10.30 \\
  1 & 3/2 & 1/2 &   47.251(4) &    0.07598 &    0.99711 &  59,70 \\
  1 & 3/2 & 3/2 &   67.746(4) &    0.06356 &    0.99798 & -42.16 \\
  {\bf 1} & {\bf 3/2} & {\bf 5/2} &   \textbf{77.202(4)} &
  \textbf{0.00000} &    \textbf{1.00000} &  \textbf{9.88} \\
\hline
 \multicolumn{7}{c}{$v=0$, $L=3$} \\
\hline
  1 & 1/2 & 7/2 & -157.925(7) &    0.98895 &   -0.14826 &    7.53 \\
  1 & 1/2 & 5/2 & -134.446(7) &    0.98227 &   -0.18749 &   -16.39 \\
  {\bf 1} & {\bf 3/2} & {\bf 3/2} &   \textbf{23.151(4)} &    \textbf{0.00000} &
  \textbf{1.00000} &    \textbf{3.87} \\
  1 & 3/2 & 5/2 &   54.118(3) &    0.18749 &    0.98227 &    6.65 \\
  1 & 3/2 & 7/2 &   80.653(4) &    0.14826 &    0.98895 &   -40.01 \\
  {\bf 1} & {\bf 3/2} & {\bf 9/2} &  \textbf{100.754(4)} &    \textbf{0.00000} &
  \textbf{1.00000} &    \textbf{16.24} \\
\hline
 \multicolumn{7}{c}{$v=1$, $L=3$} \\
\hline
  1 & 1/2 & 7/2 & -154.445(7) &    0.98944 &   -0.14497 &    7.38 \\
  1 & 1/2 & 5/2 & -131.922(7) &    0.98324 &   -0.18230 &   -15.94 \\
  {\bf 1} & {\bf 3/2} & {\bf 3/2} &   \textbf{23.914(4)} &    \textbf{0.00000} &
  \textbf{1.00000} &    \textbf{38.96} \\
  1 & 3/2 & 5/2 &   53.336(3) &    0.18230 &    0.98324 &    6.20 \\
  1 & 3/2 & 7/2 &   78.775(4) &    0.14497 &    0.98944 &   -39.85 \\
  {\bf 1} & {\bf 3/2} & {\bf 9/2} &   \textbf{98.122(4)} &    \textbf{0.00000} &
  \textbf{1.00000} &    \textbf{16.23} \\
\hline
 \multicolumn{7}{c}{$v=2$, $L=3$} \\
 1 & 1/2 & 7/2 & -151.139(7) &    0.98992 &   -0.14161 &   7.20 \\
 1 & 1/2 & 5/2 & -129.544(7) &    0.98421 &   -0.17702 & -15.42 \\
 \textbf{1} & \textbf{3/2} & \textbf{3/2} &   \textbf{24.678(4)} &    \textbf{0.00000} 
 &    \textbf{1.00000} & \textbf{38.81} \\
 1 & 3/2 & 5/2 &   52.614(3) &    0.17702 &    0.98421 &   5.72 \\
 1 & 3/2 & 7/2 &   76.992(3) &    0.14161 &    0.98992 & -39.54 \\
 \textbf{1} & \textbf{3/2} & \textbf{9/2} &   \textbf{95.605(4)} &    \textbf{0.00000} 
 &    \textbf{1.00000} &  \textbf{16.17} \\
\hline \hline\hline
 \end{tabular}
\caption{Hyperfine structure of the lower ro-vibrational states
for odd values $L=1,3$. Listed are: the quantum numbers $J$, $I$,
and $F$, the hyperfine energy $\Delta E^{\rm hfs}=\Delta
E^{(vL)IFJ}/h$ (in MHz), the amplitudes $\beta^{(vL)IFJ}_{I'F'}$
of the spin wave function, and the derivative $h^{-1}\,d\Delta
E^{\rm hfs}/dQ_d$ (in kHz~fm$^{-2}).$ The pure states are
typed in boldface.} \label{tab:odd}
\end{table}

\begin{table}
\begin{tabular}{c@{\hspace{3mm}}c@{\hspace{3mm}}c@{\hspace{3mm}}r@{\hspace{8mm}}r@{\hspace{8mm}}
 r@{\hspace{8mm}}r@{\hspace{6mm}}r@{\hspace{16mm}}}
\hline\hline
\vrule width 0pt height 10pt 
 $I$ & $F$ & $J$ & \multicolumn{1}{c}{$\Delta E^{\rm hfs}/h$} &
 \multicolumn{1}{c}{$\beta^{IFJ}_{0,1/2}$\hspace*{4mm}} &
 \multicolumn{1}{c}{$\beta^{IFJ}_{2,3/2}$\hspace*{4mm}} &
 \multicolumn{1}{c}{$\beta^{IFJ}_{2,5/2}$\hspace*{1mm}} &
 \multicolumn{1}{c}{$\left(d\Delta E^{\rm hfs}/dQ_d\right)/h$\hspace*{9mm}} \\
\hline
 \multicolumn{8}{c}{$v=0$, $L=0$} \\
\hline
  {\bf 2} & {\bf 3/2} & {\bf 3/2} & \textbf{ -213.800(11)} &    \textbf{0.00000} &
  \textbf{1.00000} &    \textbf{0.00000} &    \textbf{
  0.00000} \\
  {\bf 0} & {\bf 1/2} & {\bf 1/2} &    \textbf{0.000(8)\hspace{1pt} } &    \textbf{1.00000} &
  \textbf{0.00000} &    \textbf{0.00000} &    \textbf{0.00000} \\
  {\bf 2} & {\bf 5/2} & {\bf 5/2} &  \textbf{142.533(8)\hspace{1pt} } &    \textbf{0.00000} &
  \textbf{0.00000} &    \textbf{1.00000} &    \textbf{0.00000} \\
\hline
  \multicolumn{8}{c}{$v=1$, $L=0$} \\
\hline
  {\bf 2} & {\bf 3/2} & {\bf 3/2} & \textbf{ -209.756(11)} &    \textbf{0.00000}
   &    \textbf{1.00000} &    \textbf{0.00000} &    \textbf{0.00000} \\
  {\bf 0} & {\bf 1/2} & {\bf 1/2} &    \textbf{0.000(0)\hspace{1pt} } &    \textbf{1.00000} &
  \textbf{0.00000} &    \textbf{0.00000} &    \textbf{0.00000} \\
  {\bf 2} & {\bf 5/2} & {\bf 5/2} &  \textbf{139.837(7)\hspace{1pt} } &    \textbf{0.00000} &
  \textbf{0.00000} &    \textbf{1.00000} &    \textbf{0.00000} \\
\hline
  \multicolumn{8}{c}{$v=0$, $L=2$} \\
\hline
  2 & 3/2 & 7/2 & -226.255(11) &    0.00000 &    0.99852 &   -0.05446 &   -22.25 \\
  2 & 3/2 & 5/2 & -216.433(11) &    0.00004 &    0.99669 &   -0.08128 &    51.39 \\
  2 & 3/2 & 3/2 & -202.716(11) &   -0.00012 &    0.99664 &   -0.08194 &    5.06 \\
  2 & 3/2 & 1/2 & -190.986(11) &    0.00000 &    0.99863 &   -0.05238 &   -66.45 \\
  0 & 1/2 & 3/2 &  {-32.093(2)\hspace{1pt} } &    1.00000 &    0.00012 &    0.00006 &    0.01 \\
  0 & 1/2 & 5/2 &   {21.395(1)\hspace{1pt} } &    1.00000 &   -0.00002 &    0.00019 &   -0.02 \\
  2 & 5/2 & 1/2 &  {102.515(8)\hspace{1pt} } &    0.00000 &    0.05238 &    0.99863 &   -81.29 \\
  2 & 5/2 & 3/2 &  {117.107(7)\hspace{1pt} } &   -0.00007 &    0.08194 &    0.99664 &   -33.21 \\
  2 & 5/2 & 5/2 &  {135.572(7)\hspace{1pt} } &   -0.00019 &    0.08128 &    0.99669 &    26.03 \\
  2 & 5/2 & 7/2 &  {151.994(7)\hspace{1pt} } &    0.00000 &    0.05446 &    0.99852 &    50.39 \\
  {\bf 2} & {\bf 5/2} & {\bf 9/2} &  \textbf{159.864(8)\hspace{0.5pt} } &    \textbf{0.00000} &
  \textbf{0.00000} &    \textbf{1.00000} &   \textbf{-28.14} \\
\hline
  \multicolumn{8}{c}{$v=1$, $L=2$} \\
\hline
  2 & 3/2 & 7/2 & -221.646(11) &    0.00000 &    0.99858 &   -0.05318 &   -22.19 \\
  2 & 3/2 & 5/2 & -212.213(11) &    0.00004 &    0.99686 &   -0.07921 &    51.32 \\
  2 & 3/2 & 3/2 & -199.097(11) &   -0.00012 &    0.99682 &   -0.07963 &    4.92 \\
  2 & 3/2 & 1/2 & -187.921(11) &    0.00000 &    0.99871 &   -0.05076 &   -66.51 \\
  0 & 1/2 & 3/2 &  {-30.692(2)\hspace{1pt} } &    1.00000 &    0.00013 &    0.00007 &    0.01 \\
  0 & 1/2 & 5/2 &   {20.461(1)\hspace{1pt} } &    1.00000 &   -0.00002 &    0.00019 &   -0.02 \\
  2 & 5/2 & 1/2 &  {101.558(8)\hspace{1pt} } &    0.00000 &    0.05076 &    0.99871 &   -81.19 \\
  2 & 5/2 & 3/2 &  {115.469(7)\hspace{1pt} } &   -0.00008 &    0.07963 &    0.99682 &   -33.07 \\
  2 & 5/2 & 5/2 &  {133.118(7)\hspace{1pt} } &   -0.00019 &    0.07921 &    0.99686 &    26.07 \\
  2 & 5/2 & 7/2 &  {148.851(7)\hspace{1pt} } &    0.00000 &    0.05318 &    0.99858 &    50.33 \\
  {\bf 2} & {\bf 5/2} & {\bf 9/2} &  \textbf{156.416(7)\hspace{0.5pt} } &    \textbf{0.00000} &
  \textbf{0.00000} &    \textbf{1.00000} &   \textbf{-28.13} \\
\hline
  \multicolumn{8}{c}{$v=2$, $L=2$} \\
\hline
 2 & 3/2 & 7/2 & -217.273(11) &    0.00000 &    0.99865 &   -0.05189 & -22.05 \\
 2 & 3/2 & 5/2 & -208.218(11) &    0.00004 &    0.99702 &   -0.07710 &  51.07 \\
 2 & 3/2 & 3/2 & -195.684(11) &   -0.00012 &    0.99701 &   -0.07728 &   4.76 \\
 2 & 3/2 & 1/2 & -185.041(11) &    0.00000 &    0.99879 &   -0.04913 &  -6.63 \\
 0 & 1/2 & 3/2 &  {-29.338(2)\hspace{1pt} } &    1.00000 &    0.00013 &    0.00007 &   0.01 \\
 0 & 1/2 & 5/2 &  { 19.559(1)\hspace{1pt} } &    1.00000 &   -0.00002 &    0.00019 &  -0.02 \\
 2 & 5/2 & 1/2 &  {100.687(7)\hspace{1pt} } &    0.00000 &    0.04913 &    0.99879 &  80.80 \\
 2 & 5/2 & 3/2 &  {113.941(7)\hspace{1pt} } &   -0.00008 &    0.07728 &    0.99701 & -32.80 \\
 2 & 5/2 & 5/2 &  {130.803(7)\hspace{1pt} } &   -0.00019 &    0.07710 &    0.99702 &  26.02 \\
 2 & 5/2 & 7/2 &  {145.870(7)\hspace{1pt} } &    0.00000 &    0.05189 &    0.99865 &  50.07 \\
 {\bf 2} & {\bf 5/2} & {\bf 9/2} &  {\bf 153.139(7)\hspace{0.5pt} } & {\bf 0.00000} &
 {\bf 0.00000} & {\bf 1.00000} & {\bf -28.02} \\
\hline\hline
\end{tabular}
 \caption{
 Hyperfine structure of the lower ro-vibrational states for even
 values $L\!=\!0,2$. Listed are: the quantum numbers $J$, $I$, and $F$,
 the hyperfine energy $\Delta E^{\rm hfs}=\Delta E^{(vL)IFJ}/h$ (in
 MHz), the amplitudes $\beta^{(vL)IFJ}_{I'F'}$ of the spin wave
 function, and the derivative $h^{-1}\,d\Delta E^{\rm hfs}/dQ_d$
 (in kHz~fm$^{-2}).$ The pure states are
 typed in boldface. %
 Note that the values of $\beta_{0,1/2}^{IFJ}$, $I\!=\!0$, $F\!=\!1/2$, $L\!=\!2$ are
 strictly less than 1 but appear as $1.00000$ due to rounding to 5 significant
 digits. On the contrary, $\Delta E^{\rm hfs}$ for $I\!=\!0$,
 $F\!=\!J\!=\!1/2$ is strictly zero.
 }
 \label{tab:even}
\end{table}

 Keeping in mind the suggestion by Babb \cite{Bab1997,babb2} and Zhang et al. 
 \cite{Zhang2013,Zhang2016} to determine the deuteron electric quadrupole moment 
 $Q_d$ by means of hyperfine spectroscopy of D$_2^+$, we give in the 
 rightmost column of Tables~\ref{tab:odd} and \ref{tab:even} 
 the (numerically calculated) derivatives of the hyperfine energies 
 with respect to $Q_d$ that
 describe the sensitivity of $\Delta E^{(vL)IFJ}$ to variations of $Q_d$.
 Relevant for the determination of $Q_d$ by direct comparison of the 
 theoretical and experimental values of a specific transition frequency 
 are the differences of the
 sensitivities of upper and lower state of the transition. Some of
 these are given in Table~\ref{tab:D2p_intesities}. It can be seen
 from the largest differences ($\simeq 100\,$kHz fm$^{-2}$)
 that the experimental and theoretical uncertainties
 have to be on the order 3 Hz or less, in order to match the
 present uncertainty of $Q_d$. Such a small theoretical uncertainty cannot be 
 achieved at present. Therefore, in Sec.~\ref{sec:Q_d} we discuss an 
 alternative approach to this goal.

 The comparison of our results for the hyperfine energies $\Delta E^{(vL)IFJ}$
 with earlier calculations is illustrated in Table~\ref{tab:comparison} for the 
 $(v=0,L=1)$ state taken as representative example.
 The difference with the values of Refs.~\cite{Zhang2013,Zhang2016} 
 is in the kHz range, in agreement with the estimates 
 of the contribution from the off-diagonal matrix elements of the 
 tensor spin interaction terms which were neglected there. 
 The results agree with each other within the 
 estimated theoretical uncertainty because this contribution 
 happens to be numerically of the same order of magnitude.
 The difference with \cite{babb-psas} is larger because of the oversimplified form of the
 tensor spin interactions adopted there. 
 The difference with the experimental values of Ref.~\cite{cruse} for 
 the hyperfine shifts in the same state
 exceeds by a factor of $\sim2$ the experimental uncertainty.
 However, the authors of \cite{cruse} themselves do not rule out the possibility that the 
 overall uncertainty reported there is too optimistic. 
 
 \begin{table}[!h]
\begin{tabular}{c@{\hspace{2mm}}c@{\hspace{2mm}}c@{\hspace{7mm}}d@{\hspace{7mm}}d@{\hspace{7mm}}d}
\hline\hline
\vrule width 0pt height 10.5pt depth 0pt
 $I$ & $F$ & $J$~~ &
    \multicolumn{1}{c}{This work} &
    \multicolumn{1}{c}{\cite{Zhang2013,Zhang2016}}&
    \multicolumn{1}{c}{\cite{babb-psas}}\\
\hline
   \multicolumn{6}{c}{\vrule width 0pt height 10pt depth 0pt $(v L) = (01)$}\\
\hline
  1 &  3/2 & 5/2 &   80.624(4) &  80.623(4)  &   80.597 \\
  1 &  3/2 & 1/2 &   70.351(4) &  70.355(4)  &   70.319 \\
  1 &  3/2 & 1/2 &   47.916(4) &  47.910(3)  &   47.878 \\
  1 &  1/2 & 1/2 & -136.492(7) & -136.492(7) & -136.436 \\
  1 &  1/2 & 3/2 & -146.998(8) & -146.999(8) & -146.94  \\
\hline\hline
\end{tabular}
\caption{Comparison of the hyperfine energies $\Delta
 E^{(vL)IFJ}$ for the $(vL)=(01)$ state,
 in MHz, calculated in the present work and in
 Refs.~\cite{Zhang2013, Zhang2016} and \cite{babb-psas}.}
 \label{tab:comparison}
 \end{table}

\section{Electric quadrupole transitions}\label{sec:E2}

 \subsection{Hyperfine structure of the $E2$ spectra in
 homonuclear molecular ions}

 The evaluation of the electric quadrupole transition spectrum of
 D$_2^+$ follows closely the procedure described in details in
 \cite{Korobov2018}; we shall highlight the points that are
 specific for the D$^+_2$ molecule, and also take the opportunity to
 refine some of the definitions given there.
 The use of dimensional SI units is restored in the
 rest of the paper.

 Similar to Eqs.~(5)-(7) of \cite{Korobov2018}, we denote by
 $H_{\rm int}^{(E2)}$ the terms in the
 interaction Hamiltonian of D$^+_2$ with a monochromatic electromagnetic
 plane wave,
 which are responsible for the electric quadrupole transitions.
 We take the vector potential in the form
 ${\bf A}({\bf R},t)=(-i/\omega)\left({\bf E}_0\,e^{i({\bf k}.{\bf R}-\omega t)}-
 {\bf E}^*_0\,e^{-i({\bf k}.{\bf R}-\omega t)}\right)$, where
 ${\bf k}$ is the wave vector, $\omega=c|{\bf k}|$ is the circular frequency,
 and  ${\bf E}_0$ -- the amplitude of the oscillating electric field,

 The $E2$ transition matrix element between the initial
 $|i\rangle=|(vL)IFJJ_z\rangle$ and final
 $|f\rangle=|(v'L')I'F'J'J'_z\rangle$ hyperfine states of D$^+_2$
 is
\begin{equation}\label{eq:he2matrel}
\begin{array}{@{}l}\displaystyle
\langle \Psi^{(v'L')I'F'J'J'_z}|H_{\rm int}^{(E2)}|\Psi^{(vL)IFJJ_z}\rangle =
   \frac{i}{3c}\,\omega^{\rm NR}\,|{\mathbf E}_0|\times
\\[3mm]\displaystyle\hspace{6mm}
   \left(
      e^{-i\omega t}
         \langle \Psi^{(v'L')I'F'J'J_z'}|\, \widehat{T}^{(2)}_{ij}\,Q^{(2)}_{ij}\,|\Psi^{(vL)IFJJ_z}\rangle
      +e^{i\omega t}
         \langle \Psi^{(v'L')I'F'J'J_z'}|\, \widehat{T}^{(2)*}_{ij}\>Q^{(2)}_{ij}\,|\Psi^{(vL)IFJJ_z}\rangle
   \right),
\end{array}
\end{equation}
where $\omega^{\rm NR}=(E^{{\rm (NR)}v'L'}-E^{{\rm (NR)}vL})/\hbar$ is the transition circular frequency in the non-relativistic approximation, the asterisk denotes complex conjugation, and the tensor of the electric quadrupole transition operator is defined as
\begin{equation}\label{eq:quadd2}
Q^{(2)}_{ij} = \frac{1}{2}\sum\limits_{\alpha}Z_{\alpha}e
  \left( 3R_{\alpha i}R_{\alpha j}  - \delta_{ij}\, {\bf R}_{\alpha}^2
  \right),
\end{equation}
the summation here is over the constituents of D$_2^+$ ($\alpha=1,2$ referring to nuclei 1 and 2, and $\alpha=3$ labeling the electron), $Z_{\alpha}e$ is the corresponding electric charge. $\widehat{T}^{(2)}$ is a tensor of rank 2 with Cartesian components
\begin{equation}\label{eq:that-cartes}
\widehat{T}^{(2)}_{ij} =
   \frac{1}{2}(\hat{k}_i\hat{\epsilon}_j+\hat{k}_j\hat{\epsilon}_i),
\qquad
   \hat{\mathbf{k}}\cdot\hat{\boldsymbol{\epsilon}} = 0,
\end{equation}
 where $\hat{\mathbf{k}}=\mathbf{k}/|\mathbf{k}|$,
 and $\hat{\boldsymbol{\epsilon}}$ is a unit vector
 of polarization, $\mathbf{E}_0=\hat{\boldsymbol{\epsilon}}\,|\mathbf{E}_0|$.
 The Einstein's convention for summation over repeated
 pairs of indices of cartesian components of vectors and
 tensors is assumed in Eq.~(\ref{eq:he2matrel}) and further on.

 To switch from Cartesian to cyclic coordinates and back
 for the symmetric tensor operators of rank 2,
 we use a convention: $Q_0^{(2)} = Q_{zz}^{(2)}$, that implies
 \[
 T_{ij}^{(2)}Q_{ij}^{(2)} = \frac{3}{2}\sum_q (-1)^q T_q^{(2)} Q_{-q}^{(2)} =
   \frac{3}{2}\left(T^{(2)}\cdot Q^{(2)}\right).
 \]
 Note that, in a general case of elliptic polarization,
 the vector $\hat{\boldsymbol\epsilon}$,
 and tensor $\widehat{T}^{(2)}$ are complex.
 The matrix elements of the scalar product
 $\widehat{T}^{(2)}\cdot Q^{(2)}$ in Eq.~(\ref{eq:he2matrel})
 have the form
\begin{equation}
\begin{array}{@{}l}\displaystyle
 \langle \Psi^{(v'L')I'F'J'J_z'}|
 \widehat{T}^{(2)}_{ij} Q^{(2)}_{ij}
 |\Psi^{(vL)IFJJ_z}\rangle =
   \frac{3}{2}\sqrt{2J\!+\!1}\>\> \times
 \\[3mm]\displaystyle\hspace{15mm}
   \sum_q\widehat{T}^{(2)q}\,C_{JJ_z,2q}^{J'J'_z}\left[\,
      \sum_{I_1F_1}(-1)^{J+L+F_1}
      \left\{\begin{matrix}
         L & \!F_1\! & J \\
         J' & \!2\! & L'
      \end{matrix}\right\}
      \beta^{(vL)IFJ}_{I_1F_1} \beta^{(v'L')I'F'J'}_{I_1F_1}
   \right]
   \langle v'L'\|Q^{(2)}\|vL\rangle,
\qquad
q=J'_z\!-\!J_z,
\end{array}
\end{equation}
 and similar for the conjugate one, where
 $\langle v'L'\|Q^{(2)}\|vL\rangle$ are the reduced matrix
 elements of $Q^{(2)}$.
 The Rabi frequency for the transition
 $|i\rangle\rightarrow|f\rangle$
 is expressed in terms of these matrix elements as follows
 \begin{equation}
 \Omega_{if}=
 \frac{{\omega_{if}\,|\mathbf E}_0|}{3\hbar c}
 \langle \Psi^{(v'L')I'F'J'J_z'}|
 \widehat{T}^{(2)}_{ij} Q^{(2)}_{ij}
 |\Psi^{(vL)IFJJ_z}\rangle
 \label{eq:Rabi}
 \end{equation}
 where $\omega_{if}=(E^{(v'L')I'F'J'}-
 E^{(vL)IFJ})/\hbar$.
 Note that for general polarization 
 $\Omega_{if}$ is complex.
 The probability per unit time ${\mathcal W}_{if}$ for the
 transition $|i\rangle\rightarrow|f\rangle$, stimulated by the
 external electric field with amplitude ${\bf E}_0$, oscillating
 with frequency $\omega$ and propagating along ${\bf k}$,
 may be expressed in terms of the Rabi frequency as follows:
 ${\mathcal W}_{if}=2\pi(\delta(\omega-\omega_{if})+
 \delta(\omega+\omega_{if}))|\Omega_{if}|^2$. We shall put it in
 the factorized form used in Eq.~(18) of \cite{Korobov2018}:
 \begin{equation}
 {\mathcal W}_{if}={\mathcal W}^{\rm NR}(v'L';vL)\,
 {\mathcal W}^{\rm hfs}((v'L')I'F'J';(vL)IFJ)\,
 {\mathcal W}^{\rm pol}(J'J'_z;J_z).
 \label{eq:fact}
 \end{equation}
 The first factor, ${\mathcal W}^{\rm NR}(v'L';vL)$, is the rate of stimulated
 $E2$ transitions in D$^+_2$ in the non-relativistic (spinless)
 approximation, averaged over the initial and summed over the final
 angular momentum projections $L_z,L'_z$
\begin{equation}\label{eq:W^nr}
\mathcal{W}^{\rm NR}(v'L';vL) =
   \frac{\pi\omega_{if}^2}{\varepsilon_0c^3\hbar^2}\frac{1}{15(2L+1)}
   \left|\langle v'L'\|Q^{(2)}\|vL\rangle\right|^2\,\bar{\mathcal I},
\qquad
\bar{\mathcal I} = \int d\omega\,\mathcal{I}(\omega)g_{if}(\omega),
\end{equation}
 where ${\mathcal I}(\omega)$ is the spectral density of the
 external (laser) field energy flux, and $g_{if}(\omega)$
 is the transition line spectral profile.
 The factor 
\begin{equation}\label{eq:W^hfs}
\begin{array}{@{}l}\displaystyle
{\mathcal W}^{\rm hfs}((v'L')I'F'J';(vL)IFJ) =
   (2L\!+\!1)(2J'\!+\!1)
   \left(
      \sum_{I_1F_1}\beta^{(v'L')I'F'J'}_{I_1F_1}\beta^{(vL)IFJ}_{I_1F_1}
      (-1)^{J+F_1}
      \left\{\begin{matrix}
         L & \!F_1\! & J \\
         J' & \!2\! & L'
      \end{matrix}\right\}
   \right)^2,
\end{array}
\end{equation}
 describes the intensity of the individual hyperfine component
 $IFJ\rightarrow I'F'J'$ of the transition line
 $(vL)\to(v'L')$.
 Because of the weak mixing of the pure states
 in Eq.~(\ref{eq:betaexpans}) (see
 Tables~\ref{tab:odd},\ref{tab:even})
 a good approximation for the intensity of the strong (favored)
 transitions is to assume that
 $\beta^{(vL)IFJ}_{I'F'}\approx\delta_{I'I}\delta_{F'F}$ that leads to
 \begin{equation}\label{eq:W^hfs-strong}
 \begin{array}{@{}l}\displaystyle
 {\mathcal W}^{\rm hfs}((v'L')IFJ';(vL)IFJ) \approx
   (2L\!+\!1)(2J'\!+\!1)
      \left\{\begin{matrix}
         L & \!F\! & J \\
         J' & \!2\! & L'
      \end{matrix}\right\}^2.
 \end{array}
 \end{equation}
 The weak (unfavored) transitions are forbidden in this
 approximation.

 Finally
 \begin{equation}\label{W^pol}
 {\mathcal W}^{\rm pol}(J'J'_z;JJ_z) =
   \frac{15(2J+1)}{2J'+1}
   \left(C_{JJ_z,2q}^{J'J'_z}\right)^2
   \left|\widehat{T}^{(2)}_q\right|^2,
 \qquad
 q=J'_z-J_z,
 \end{equation}
 is related to the intensity of
 the Zeeman components of a hyperfine transition line with
 different values of the magnetic quantum numbers $J_z, J'_z$.
 ${\mathcal W}^{\rm pol}(J'J'_z;JJ_z)$ is normalized with the condition
\begin{equation}\label{eq:normW^pol}
\frac{1}{2J\!+\!1}\sum_{J_z;J'_z}{\mathcal W}^{\rm
pol}(J'J'_z,JJ_z)=1.
\end{equation}
 Note that compared with Eqs.~(20)-(21) of Ref.~\cite{Korobov2018},
 the factor $(2J\!+\!1)$ has now been moved from
 ${\mathcal W}^{\rm hfs}$ to ${\mathcal W}^{\rm pol}$.
 In absence of external magnetic field or in case of
 spectral resolution insufficient to distinguish the
 Zeeman components, the rate ${\mathcal W}_1$ of excitation
 of an individual
 hyperfine component $J_z$ to any of the Zeeman states $J'_z$
 is independent of $J_z$,
 and using Eqs.~Eq.~(\ref{eq:fact}),(\ref{eq:normW^pol}) is reduced to
 ${\mathcal W}_{1}={\mathcal W}^{\rm NR}(v'L';vL)\,
 {\mathcal W}^{\rm hfs}((v'L')I'F'J';(vL)IFJ)$.
 Similarly, the product
 ${\mathcal W}^{\rm hfs}((v'L')I'F'J';(vL)IFJ)\
 {\mathcal W}^{\rm pol}(J'J'_z;JJ_z)$ satisfies the normalization
 condition
 \begin{equation}
 \sum_{I'F'J'J'_z}\frac{1}{n^{\rm hfs}(vL)}\sum_{IFJJ_z}
  {\mathcal W}^{\rm hfs}((v'L')I'F'J';(vL)IFJ)\,{\mathcal W}^{\rm pol}(J'J'_z;JJ_z)=1,
 \end{equation}
 where the sum is over the allowed values of $I,I'$ with the same
 parity as $L$,
 and $n^{\rm hfs}(vL)$ is the number of states in the considered hyperfine manifold of the $(vL)$ state:
 \begin{equation}
 n^{\rm hfs}(vL)=
 \begin{cases}
 6(2L+1)\text{ for odd }L,\\
 12(2L+1)\text{ for even }L.
 \end{cases}
 \end{equation}
 In case the hyperfine structure of the $E2$ transition line
 is not resolved,
 the excitation rate ${\mathcal W}_2$
 from any of the $n^{\rm hfs}(vL)$ initial
 states to all final states is found by summing
 ${\mathcal W}_{if}$ of Eq.~(\ref{eq:fact}) over all final states and
 averaging over all initial states. The result is
 ${\mathcal W}_2={\mathcal W}^{\rm NR}(v'L';vL)$.

 \subsection{Laser polarization effects on the Zeeman structure of $E2$ spectra}

 \begin{figure}[t]
 \includegraphics[width=0.6\textwidth]{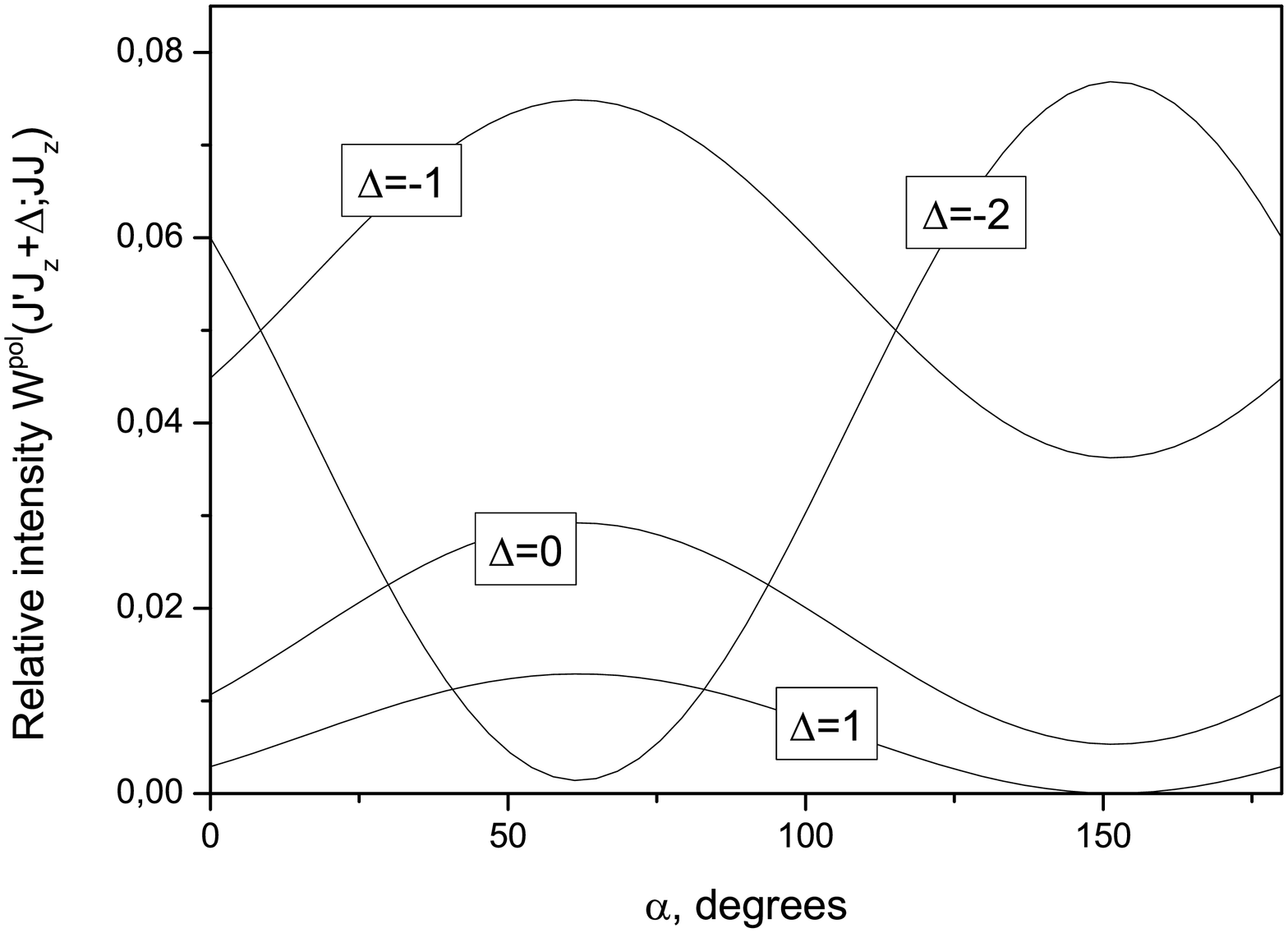}
 \caption{Relative intensities ${\mathcal W}^{\rm pol}(J'J'_z;JJ_z)$
 of the Zeeman components
 $(JJ_z)=(\frac{1}{2}\frac{1}{2})\rightarrow(J'J'_z)=(\frac{3}{2}\frac{1}{2}+\Delta)$,
 $\Delta=-2,\ldots,1$ as function of $\alpha$ for randomly selected
 fixed values of the angles  $\beta=72^{\circ}, \theta=56^{\circ}, \varphi=51^{\circ}$
 (cf.~ Eq.~(\ref{eq:TTTgen}))}.
 \label{fig:wpol1}
 \end{figure}

 \begin{figure}[t]
 \includegraphics[width=0.6\textwidth]{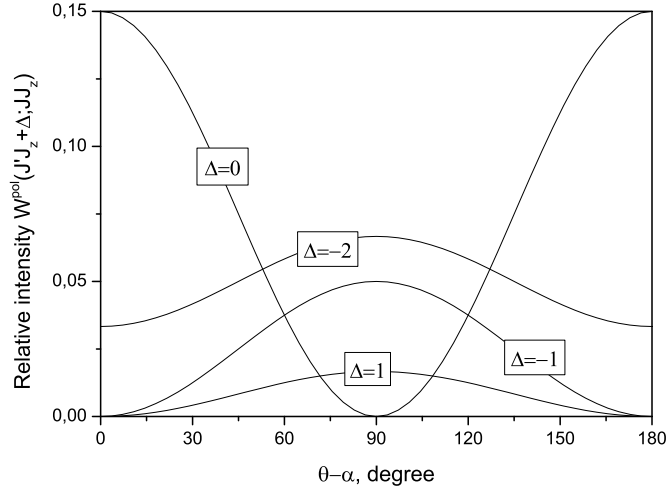}
 \caption{Relative intensities ${\mathcal W}^{\rm pol}(J'J'_z;JJ_z)$
 of the Zeeman components $(JJ_z)=(\frac{1}{2}\frac{1}{2})\rightarrow(J'J'_z)=
 (\frac{3}{2}\frac{1}{2}+\Delta)$, $\Delta=-2,\ldots,1$,
 stimulated with linearly polarized laser light, as function of $\theta-\alpha$ for $\beta=45^{\circ}$ (cf.~Eq.~(\ref{eq:TTTlin})).}
 \label{fig:wpol2}
 \end{figure}

 \begin{table}[t]
\begin{tabular}{@{\hspace{2mm}}c@{\hspace{4mm}}c@{\hspace{4mm}}r@{\hspace{10mm}}c@{\hspace{8mm}}c@{\hspace{6mm}}c@{\hspace{2mm}}}
 \hline\hline
 \vrule width 0pt height 11pt depth 5pt & & &
 \multicolumn{2}{c}{$A_{v'L'\rightarrow vL}$\ (s$^{-1}$)} \\
 \cline{5-6}
 \vrule width 0pt height 12.5pt
    $(vL)$ & $(v'L')$
  & $\Delta E^{\rm NR}/hc\ (\mbox{cm}^{-1})$\hspace{-6mm}
  & $\bigl|\langle v'L'\|Q^{(2)}\|vL\rangle\bigr|/ea_0^2$
  & This work
  & \multicolumn{1}{c}{\hspace*{0mm}Pilon \cite{Pilon2013}} \\
\hline
\vrule width 0pt height 10pt 
 $(00)$ & $(02)$ &     88.053 &   1.608226 &  0.30665[$-$12] &  0.30665[$-$12] \\
 $(00)$ & $(12)$ &   1661.833 &   0.267274 &  0.20281[$-$07] &  0.20281[$-$07] \\
 $(00)$ & $(22)$ &   3171.009 &   0.019403 &  0.27036[$-$08] &  0.27036[$-$08] \\
 $(00)$ & $(32)$ &   4617.211 &   0.002497 &  0.29299[$-$09] &  0.29298[$-$09] \\
 $(00)$ & $(42)$ &   6001.881 &   0.000451 &  0.35543[$-$10] \\
 $(00)$ & $(62)$ &   8591.673 &   0.000027 &  0.77727[$-$12] \\
 $(01)$ & $(11)$ &   1575.973 &   0.311496 &  0.35215[$-$07] &  0.35215[$-$07] \\
 $(01)$ & $(21)$ &   3087.284 &   0.019765 &  0.40905[$-$08] &  0.40905[$-$08] \\
 $(01)$ & $(31)$ &   4535.575 &   0.002284 &  0.37384[$-$09] &  0.37382[$-$09] \\
 $(01)$ & $(41)$ &   5922.282 &   0.000367 &  0.36679[$-$10] \\
 $(02)$ & $(12)$ &   1573.782 &   0.340448 &  0.25064[$-$07] &  0.25064[$-$07] \\
 $(02)$ & $(22)$ &   3082.958 &   0.021629 &  0.29183[$-$08] &  0.29182[$-$08] \\
 $(02)$ & $(42)$ &   5913.830 &   0.000403 &  0.26277[$-$10] \\
 $(03)$ & $(11)$ &   1429.607 &   0.420823 &  0.39479[$-$07] &  0.39479[$-$07] \\
 $(04)$ & $(12)$ &   1369.684 &   0.522608 &  0.29490[$-$07] \\
 $(04)$ & $(24)$ &   3067.885 &   0.027948 &  0.26415[$-$08] \\
 $(04)$ & $(32)$ &   4325.062 &   0.001717 &  0.99945[$-$10] &  0.99939[$-$10] \\
 $(06)$ & $(56)$ &   7145.142 &   0.000129 &  0.26793[$-$11] \\
 $(11)$ & $(13)$ &    140.894 &   2.379256 &  0.50287[$-$11] &  0.50287[$-$11] \\
 $(11)$ & $(33)$ &   3089.955 &   0.048979 &  0.10811[$-$07] &  0.10811[$-$07] \\
 $(20)$ & $(42)$ &   2912.431 &   0.052069 &  0.12725[$-$07] \\
 $(32)$ & $(64)$ &   4135.286 &   0.024586 &  0.90959[$-$08] \\
 $(52)$ & $(54)$ &    167.679 &   4.033412 &  0.26834[$-$10] \\
\hline\hline
\end{tabular}
\caption{Nonrelativistic energies, $\Delta E^{\rm NR}=E^{{\rm
(NR)}v'L'}\!-\!E^{{\rm (NR)}vL}$, reduced matrix elements,
$|\langle v'L'||Q^{(2)}||vL\rangle\|$, of the electrical
quadrupole moments, $Q^{(2)}$, Eq.~(\ref{eq:quadd2}), Einstein
coefficients, $A_{v'L'\rightarrow vL}$ for selected $E2$
transitions between the ro-vibrational states $|vL\rangle$ and
$|v'L'\rangle$ of ${\rm D_2^+}$, and, for comparison, the results
of H.O.~Pilon \cite{Pilon2013} when available.}
\label{tab:redElemD2plus}
\end{table}

 Recent progress in the precision spectroscopy of the HD$^+$ molecular ion 
 \cite{Alighanbari-2020} has made it possible to
 resolve the Zeeman structure of laser induced $E1$ transition
 lines. This allowed to quantitatively study of the Zeeman and Stark 
 shifts of the spectral lines, to reduce the related systematic uncertainties,
 and to determine the electron-to-proton mass ratio with improved accuracy.
 Anticipating the future precision spectroscopy of the D$_2^+$ ion, 
 we consider here some general characteristics of the Zeeman structure 
 of laser-induces electric quadrupole transitions and reveal effects 
 of the polarization of the stimulating laser light 
 that appear only in higher multipolarity spectra.
 These effects are described
 by the factor ${\mathcal W}^{\rm pol}(J'J'_z;JJ_z)$ in Eq.~(\ref{eq:fact})
 and impact the intensity and not the
 frequency of the $E2$ transition lines;
 the calculations of the Zeeman shift of the transition
 frequencies will be published elsewhere.

 The cyclic components of the rank-2 irreducible tensor $\widehat{T}^{(2)}$
 (\ref{eq:that-cartes}) are expressed in terms of the cartesian
 components as follows:
\begin{equation}\label{eq:that-cyc}
\begin{array}{@{}l}\displaystyle
\widehat{T}^{(2)\pm2} =
   \sqrt{\frac{1}{6}}
   \left(
      \hat{k}_x\hat{\epsilon}_x-\hat{k}_y\hat{\epsilon}_y
      \mp i\left(\hat{k}_x\hat{\epsilon}_y+\hat{k}_y\hat{\epsilon}_x\right)
   \right)
\\[3mm]\displaystyle
\widehat{T}^{(2)\pm1} =
   \sqrt{\frac{1}{6}}
   \left(
      \mp\left(\hat{k}_x\hat{\epsilon}_z+\hat{k}_z\hat{\epsilon}_x\right)
      +i\left(\hat{k}_z\hat{\epsilon}_y+\hat{k}_y\hat{\epsilon}_z\right)
   \right)
\\[3mm]\displaystyle
\widehat{T}^{(2)0} =
   \frac{1}{3}\left(2\hat{k}_z\hat{\epsilon}_z-\hat{k}_x\hat{\epsilon}_x-\hat{k}_y\hat{\epsilon}_y\right) =
   \hat{k}_z\hat{\epsilon}_z.
\end{array}
\end{equation}
 We have updated the normalization of these components
 as compared with Ref.~\cite{Korobov2018}, but have kept unchanged
 the parametrization of  the complex unit
 vector $\hat{\boldsymbol{\epsilon}}={\mathbf E}_0/|{\mathbf E}_0|$
 pointing along the electric
 field amplitude ${\mathbf E}_0$.
 As in \cite{Korobov2018}, we denote by $K$ the lab reference frame
 with $z$-axis along the external magnetic
 field ${\mathbf B}$, by $K'$ a reference frame
 with $z$-axis along $\hat{\bf k}$,
 and take the cartesian coordinates $(\epsilon'_x,\epsilon'_y,\epsilon'_z)$ of
 $\hat{\boldsymbol{\epsilon}}$ in
 $K'$ to be $(\cos\theta, \sin\theta\,e^{i\varphi},0)$.
 Linear polarization of the incident light is described by
 $\varphi=0$; left/right circular polarization -- by
 $\varphi=\pm\pi/2, \theta=\pi/4$; all other combinations
 correspond to general elliptic polarization. Let
 $(\alpha,\beta,\gamma)$ be the Euler angles of the rotation that
 transforms $K$ into $K'$, and denote by $M(\alpha,\beta,\gamma)$
 the matrix relating the cartesian coordinates $(a_x,a_y,a_z)$ and
 $(a'_x,a'_y,a'_z)$ of an arbitrary vector ${\mathbf a}$ in $K$ and
 $K'$, respectively: $a_i=\sum_j M_{ij}(\alpha,\beta,\gamma)\,a'_j$.
 (To avoid mismatch of $M$ with $M^{-1}$,
 note that, e.g. $M_{xz}=-\sin\beta\,\cos\gamma$.)
 In this way, the absolute values of the components of $\widehat{T}$ in the lab frame $K$,
 appearing in Eq.~(\ref{W^pol})
 are parametrized with the {\em four} angles
 $\alpha,\beta,\theta$, and $\varphi$ (the dependence on $\gamma$
 being cancelled):
 \begin{eqnarray}
 &\left|\widehat{T}^{(2)\pm2}\right|^2=&
 -\frac{1}{12}\sin^4\beta\,(1+\cos2\alpha\,\cos2\theta+\sin2\alpha\,
 \sin2\theta\,\cos\varphi)+
 \nonumber
 \\
 &&
 \frac{1}{6}\sin^2\beta(1\pm\cos\beta\,\sin2\theta\,\sin\varphi)
 \nonumber
 \\
 &\left|\widehat{T}^{(2)\pm1}\right|^2=&\frac{1}{12}+
 \frac{1}{24}\cos2\beta\,(1-\cos2\alpha\,\cos2\theta)+
 \frac{1}{24}\cos4\beta\,(1+\cos2\alpha\,\cos2\theta)-
 \nonumber
 \\
 &&\frac{1}{12}(1+2\cos2\beta)\sin^2\beta\,\sin2\alpha\,\sin2\theta\,\cos\varphi\pm
 \frac{1}{6}\cos\beta\,\cos2\beta\,\sin2\theta\,\sin\varphi
 \nonumber
 \\
 &\left|\widehat{T}^{(2)0}\right|^2=&\frac{1}{8}\sin^2 2\beta\,
 (1+\cos2\alpha\,\cos2\theta+\sin2\alpha\,\sin2\theta\,\cos\varphi)
 \label{eq:TTTgen}
 \end{eqnarray}
 This leads, for linear polarization ($\varphi=0$), to
 \begin{eqnarray}
 &\left|\widehat{T}^{(2)\pm2}_{\rm lin}\right|^2=&
 \frac{1}{6}\sin^2\beta\,(1-\sin^2\beta\,\cos^2(\theta-\alpha)) \nonumber
 \\
 &\left|\widehat{T}^{(2)\pm1}_{\rm lin}\right|^2=&\frac{1}{12}\,
 (1+\cos4\beta\,\cos^2(\theta-\alpha)+\cos2\beta\,\sin^2(\theta-\alpha))
 \nonumber
 \\
 &\left|\widehat{T}^{(2)0}_{\rm lin}\right|^2=&\frac{1}{4}\sin^2 2\beta\,
 \cos^2(\theta-\alpha),
 \label{eq:TTTlin}
 \end{eqnarray}
 and for left circular polarization $(\theta=\pi/4,
 \varphi=\pi/2)$
 \begin{eqnarray}
 &\left|\widehat{T}^{(2)\pm2}_{\rm left}\right|^2=&
 \frac{1}{3}\sin^2\beta\,
 \left(
 \begin{matrix}
 \cos^4\frac{\beta}{2}\\
 \sin^4\frac{\beta}{2}
 \end{matrix}
 \right)
 \nonumber
 \\
 &\left|\widehat{T}^{(2)\pm1}_{\rm left}\right|^2=&\frac{1}{3}\,
 (1\mp2\cos\beta)^2
 \left(
 \begin{matrix}
 \cos^4\frac{\beta}{2}\\
 \sin^4\frac{\beta}{2}
 \end{matrix}
 \right)
 \nonumber
 \\
 &\left|\widehat{T}^{(2)0}_{\rm left}\right|^2=&\frac{1}{8}\sin^2 2\beta.
 \label{eq:TTTcir}
 \end{eqnarray}
  For right circular polarization, described by $\theta=\pi/4$,
  $\varphi=-\pi/2$, the values of $|\widehat{T}^{(2)q}_{\rm right}|^2$ are
  obtained from the above expressions with the substitution
  $\left|\widehat{T}^{(2)q}_{\rm right}\right|^2=
  \left|\widehat{T}^{(2)-q}_{\rm left}\right|^2$.

\begin{figure}[t]
\includegraphics[width=0.6\textwidth]{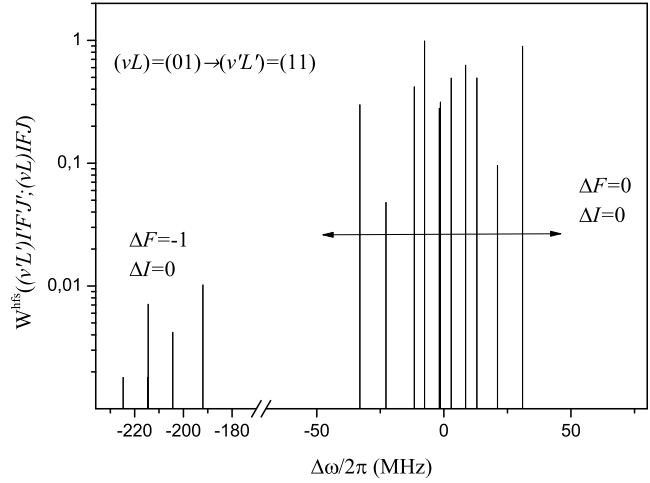}
 \caption{Hyperfine structure of the $|01\rangle\rightarrow|11\rangle$
 $E2$ vibrational transition line.
 The intensity of the hyperfine components ${\mathcal W}^{\rm hfs}$
 is plotted against the
 laser frequency detuning $\Delta\omega/2\pi$ from the spin-averaged transition
 frequency.
 The spectrum is dominated by the ``strong''
 (favored) components with $\Delta F=\Delta I=0$, while the intensity of
 transitions between hyperfine  states with different $F$ or $I$
 is suppressed by two orders of magnitude.
 The spectrum also includes a number of ``weak'' (unfavored)
 transitions with $\Delta F=+1$, not shown on the plot.}
 \label{fig:whfs}
\end{figure}

 One might expect the intensity of the Zeeman components of the
 $E2$-transition spectrum to depend ---
 as in $E1$ transitions --- on three parameters only: one
 related to the laser polarization, and two more
 describing the mutual orientation of the external magnetic field
 ${\bf B}$ and the unit vectors $\hat{\mathbf{k}}$ and $\hat{\boldsymbol{\epsilon}}$.
 In fact,  Eqs.~(\ref{eq:TTTlin}) and (\ref{eq:TTTcir})
 show that this is the case for circular and linear polarization
 only (when the difference $\theta-\alpha$ appears as a single
 parameter) while in the general case
 ${\mathcal W}^{\rm pol}(J'J'_z;JJ_z)$ depends substantially
 on both angles $\alpha$ and $\theta$.
 As an illustration, on Fig.~\ref{fig:wpol1} are plotted
 the relative intensities of the Zeeman components
 $|(vL)IFJ,J_z\rangle=|(00)0\frac{1}{2}\frac{1}{2},\frac{1}{2}\rangle
 \rightarrow|(12)0\frac{1}{2}\frac{3}{2},\frac{1}{2}+\Delta\rangle,
 \Delta=-2,-1,0,1$ as functions of $\alpha$ for the randomly selected
 values $\beta=72^{\circ}, \theta=56^{\circ}, \varphi=51^{\circ}$.
 The plot shows that the measurement of one or other individual Zeeman
 component of $E2$ transition lines may be substantially enhanced
 with appropriate optimization of the set-up geometry using
 Eqs.~(\ref{eq:TTTgen})-(\ref{eq:TTTcir}).
 The rather sharp
 dependence of the intensity of the individual Zeeman components
 on $\theta-\alpha$ for linear polarization and
 fixed value of the angle $\beta$ between
 ${\bf B}$ and the laser propagation direction (cf.
 Eq.~(\ref{eq:TTTlin}))
 is shown in Fig.~\ref{fig:wpol2}.

 \subsection{Numerical results}

\begin{table}[t]
\begin{tabular}{r@{\hspace{5mm}}c@{\hspace{5mm}}c@{\hspace{4mm}}c@{\hspace{4mm}}c@{\hspace{13mm}}r@{\hspace{12mm}}c
 @{\hspace{12mm}}r@{\hspace{18mm}}}
 \vspace{-5mm}\\
\hline\hline
\vrule width 0pt height 9.5pt depth 1.0pt
 $i$ & $I$ & $F$ & $J$ & $J'$ &
    \multicolumn{1}{c}{\hspace*{-10mm}$\Delta E^{\rm hfs}$ (MHz)} &
    \multicolumn{1}{c}{\hspace*{-10mm}${\mathcal W}^{\rm hfs}$} &
    \multicolumn{1}{c}{$(d\Delta E^{\rm hfs}/dQ_d)/h$\ (kHz\,fm$^{-2}$)}\\
    \hline
 \multicolumn{6}{c}{\vrule width 0pt height 10pt $(vL)=( 00)\to(v'L')=( 02)$}\\
 \hline
1& 2 & 5/2 &  5/2 &  1/2 &    -40.018(11) &  0.06648 &    -81.29 \\
2& 0 & 1/2 &  1/2 &  3/2 &    -32.093(2)\phantom{0} &  0.40000 &      0.01 \\
3& 2 & 5/2 &  5/2 &  3/2 &    -25.426(11) &  0.13244 &    -33.21 \\
4& 2 & 3/2 &  3/2 &  7/2 &    -12.456(16) &  0.39882 &    -22.25 \\
5& 2 & 5/2 &  5/2 &  5/2 &     -6.961(11) &  0.19868 &     26.03 \\
6& 2 & 3/2 &  3/2 &  5/2 &     -2.633(16) &  0.29802 &     51.39 \\
7& 2 & 5/2 &  5/2 &  7/2 &      9.461(11) &  0.26588 &     50.39 \\
8& 2 & 3/2 &  3/2 &  3/2 &     11.084(16) &  0.19866 &      5.06 \\
 {\bf 9}& {\bf 2} & {\bf 5/2} &  {\bf 5/2} &  {\bf 9/2} &  {\bf 17.331(11)} &  {\bf 0.33333} & {\bf -28.14} \\
 {\bf 10}& {\bf 0} & {\bf 1/2} &  {\bf 1/2} &  {\bf 5/2} &  {\bf 21.395(1)\phantom{0}} &  {\bf 0.60000} & {\bf -0.02} \\
11& 2 & 3/2 &  3/2 &  1/2 &     22.814(16) &  0.09973 &    -66.45 \\
\hline
 \multicolumn{6}{c}{\vrule width 0pt height 10pt $(vL)=( 00)\to(v'L')=( 12)$}\\
 \hline
1& 2 & 5/2 &  5/2 &  1/2 &    -40.975(11) &  0.06649 &    -81.20 \\
2& 0 & 1/2 &  1/2 &  3/2 &    -30.692(2)\phantom{0} &  0.40000 &      0.01 \\
3& 2 & 5/2 &  5/2 &  3/2 &    -27.064(11) &  0.13249 &    -33.07 \\
4& 2 & 5/2 &  5/2 &  5/2 &     -9.415(11) &  0.19875 &     26.07 \\
5& 2 & 3/2 &  3/2 &  7/2 &     -7.846(16) &  0.39886 &    -22.19 \\
6& 2 & 3/2 &  3/2 &  5/2 &      1.587(16) &  0.29812 &     51.32 \\
7& 2 & 5/2 &  5/2 &  7/2 &      6.318(11) &  0.26591 &     50.33 \\
{\bf 8} &{\bf 2} & {\bf 5/2} &  {\bf 5/2} &  {\bf 9/2} &     {\bf 13.883(11)} &
 {\bf 0.33333} & {\bf -28.13} \\
9& 2 & 3/2 &  3/2 &  3/2 &     14.702(16) &  0.19873 &      4.92 \\
{\bf 10} &{\bf 0} & {\bf 1/2} &  {\bf 1/2} &  {\bf 5/2} &     {\bf 20.461(1)\phantom{0}} &
 {\bf 0.60000} &  {\bf -0.02} \\
11& 2 & 3/2 &  3/2 &  1/2 &     25.879(16) &  0.09974 &    -66.51 \\
 \hline                          
 \multicolumn{6}{c}{\vrule width 0pt height 10pt $(vL)=( 00)\to(v'L')=( 22)$}\\
 \hline
1& 2 & 5/2 &  5/2 &  1/2 &    -41.846(11) &  0.06651 &    -80.80 \\
2& 0 & 1/2 &  1/2 &  3/2 &    -29.338(2)\phantom{0} &  0.40000 &      0.01 \\
3& 2 & 5/2 &  5/2 &  3/2 &    -28.592(10) &  0.13254 &    -32.80 \\
4& 2 & 5/2 &  5/2 &  5/2 &    -11.730(10) &  0.19881 &     26.02 \\
5& 2 & 3/2 &  3/2 &  7/2 &     -3.474(16) &  0.39892 &    -22.05 \\
6& 2 & 5/2 &  5/2 &  7/2 &      3.337(10) &  0.26595 &     50.07 \\
7& 2 & 3/2 &  3/2 &  5/2 &      5.582(16) &  0.29821 &     51.07 \\
{\bf 8}& {\bf 2} & {\bf 5/2} &  {\bf 5/2} &  {\bf 9/2} &     
 {\bf 10.606(11)} &  {\bf 0.33333} & {\bf -28.02} \\
9& 2 & 3/2 &  3/2 &  3/2 &     18.115(16) &  0.19881 &      4.76 \\
{\bf 10}& {\bf 0} & {\bf 1/2} &  {\bf 1/2} &  {\bf 5/2} &     {\bf 19.559(1)\phantom{0}} 
 &  {\bf 0.60000} & {\bf -0.02} \\
11& 2 & 3/2 &  3/2 &  1/2 &     28.759(16) &  0.09976 &    -66.32 \\
 \hline
 \multicolumn{6}{c}{\vrule width 0pt height 10pt $(vL)=( 01)\to(v'L')=( 11)$}\\
 \hline
1& 1 & 3/2 &  5/2 &  1/2 &    -33.061(5)\phantom{0} &  0.29815 &     50.33 \\
2& 1 & 3/2 &  3/2 &  1/2 &    -22.787(5)\phantom{0} &  0.04783 &    102.72 \\
3& 1 & 3/2 &  5/2 &  3/2 &    -11.612(5)\phantom{0} &  0.41821 &    -52.31 \\
4& 1 & 1/2 &  1/2 &  3/2 &     -7.593(11) &  0.98593 &     13.69 \\
{\bf 5}& {\bf 1} & {\bf 3/2} &  {\bf 5/2} &  {\bf 5/2} &     {\bf -1.754(5)\phantom{0}} 
 &  {\bf 0.28000} &      {\bf -0.00} \\
6& 1 & 3/2 &  3/2 &  3/2 &     -1.339(5)\phantom{0} &  0.31375 &      0.08 \\
7& 1 & 1/2 &  3/2 &  3/2 &      2.913(11) &  0.49218 &     -0.06 \\
8& 1 & 3/2 &  3/2 &  5/2 &      8.519(5)\phantom{0} &  0.62718 &     52.39 \\
9& 1 & 1/2 &  3/2 &  1/2 &     12.973(10) &  0.49305 &    -13.44 \\
10& 1 & 3/2 &  1/2 &  3/2 &     21.096(5)\phantom{0} &  0.09564 &   -102.97 \\
11& 1 & 3/2 &  1/2 &  5/2 &     30.954(5)\phantom{0} &  0.89412 &    -50.66 \\
                \hline
                \hline
\end{tabular}
 \caption{The hyperfine shifts $\Delta E^{\rm hfs}/h=
 (\Delta E^{(v'L')IFJ'}-\Delta E^{(vL)IFJ})/h$,
 in MHz, the relative intensities ${\mathcal W}^{\rm hfs}={\mathcal W}^{\rm
 hfs}((v'L')IFJ';(vL)IFJ)$,
 and the derivative $(d\Delta E^{\rm hfs}/dQ_d)/h$\ 
 for the "strong" (favored) components in the hyperfine spectrum
 of two $E2$ transitions in D$_2^+$. The transitions between
 stretched states are shown in boldface.}
 \label{tab:D2p_intesities}
\end{table}

 The rate ${\cal W}^{\rm NR}(v'L';vL)$ of laser stimulated $E2$ transitions between the
 ro-vibrational states $|vL\rangle$ and $|v'L'\rangle$ of the
 molecular ion D$^+_2$ are expressed in Eq.~(\ref{eq:W^nr}) in terms of
 the reduced matrix elements $\langle v'L'\|Q^{(2)}\|vL\rangle$ of
 the electric quadrupole moment of D$^+_2$ between these states.
 In the present work the reduced matrix elements were
 evaluated using the non-relativistic wave
 functions of D$^+_2$ calculated in the variational approach of
 Ref.~\cite{KorobovVarMethod}. The numerical values needed for
 the evaluation of the rate of $E2$ transitions between
 a few selected ro-vibrational states are given in
 Table~\ref{tab:redElemD2plus}. For these transitions the table also
 lists the values of the Einstein's coefficients $A_{v'L'\rightarrow vL}$,
 which are related to the reduced matrix elements by
 \begin{equation}
 A_{v'L'\rightarrow vL}/t_0^{-1} =
 \frac{\alpha^5}{15(2L+1)}((E^{{\rm NR}v'L'}-E^{{\rm NR}vL})/
 \mathcal{E}_0)^5(\langle v' L'\|Q^{(2)}\| v L
 \rangle/ea_0^2)^2,
 \end{equation}
 where $a_0$, $t_0=a_0/\alpha c$, and $\mathcal{E}_0=2{\rm Ry}$ are the
 atomic units of length, time, and energy. The comparison with the values
 calculated by  H. O. Pilon \cite{Pilon2013} with different methods for a partly
 overlapping selection of transitions
 shows good agreement for the lower excited states, and
 indications of possible discrepancy of the order of $10^{-4}$ for
 the higher vibrational excitations.
 Juxtaposition with the analogous Table II of Ref.~\cite{Korobov2018} shows that
 the rates of spontaneous transitions in
 D$^+_2$ are suppressed in comparison with H$_2^+$. This is mainly due to the smaller
 rotational and vibrational excitation energies, related to the
 larger nuclear mass.
 In addition to Table~\ref{tab:redElemD2plus}, the electronic supplement
 \cite{suppl} gives the list of the reduced
 matrix elements for all $E2$ transitions between the
 ro-vibrational states with $v\le10$ and $L\le4$.

 Further on, using the values of the amplitudes
 $\beta^{(vL)IFJ}_{I'F'}$, obtained by diagonalization of the
 effective spin Hamiltonian matrix (\ref{eq:eigprob}), we
 calculated the coefficients ${\cal W}^{\rm hfs}((v'L')I'F'J';(vL)IFJ)$,
 in the expression (\ref{eq:fact}) for the rate of the individual hyperfine
 components. Fig.~\ref{fig:whfs} illustrates the hyperfine
 structure of the $E2$ transition
 $|01\rangle\rightarrow|11\rangle$ and is representative also for
 other ro-vibrational transitions.

 The spectrum consists of ``strong'' (favored) components between hyperfine
 states with the same values of the quantum numbers $F$ and $I$
 spread over a range up to $\pm50$ MHz around the center of
 gravity of the hyperfine manifold, and ``weak'' components
 between states with $\Delta F\ne0$ or $\Delta I\ne0$ at a
 distance of a few hundred MHz. The weak components are suppressed
 due to the relatively weak mixing of $F$ and $I$ in the
 eigenstates of the effective spin Hamiltonian matrix.
 Compared to H$^+_2$, however, the suppression in D$^+_2$ is less
 pronounced; the reason is that, because of the smaller nuclear
 magnetic moment of the deuteron, the contact spin-spin
 interactions dominate to a lesser extent thus leaving room for
 more $F$ and $I$ mixing.

 Table~\ref{tab:D2p_intesities} lists the details of the
 ``strong'' (favored) hyperfine components of four selected $E2$ transition
 lines: one rotational transition, two fundamental vibrational
 transitions, and one vibrational overtone transition.
 The electronic supplement \cite{suppl} includes a table of the hyperfine
 structure of all $E2$ transition lines between the ro-vibrational
 states with $v\le10$ and $L\le4$.
 
 \subsection{Determining $Q_d$ by the composite frequency method}
 \label{sec:Q_d}
 
 The currently available most accurate values of the deuteron electric 
 quadrupole moment $Q_d$ have been obtained by combining the experimental results 
 about the tensor interaction constant (traditionally denoted by $d$) 
 in the $L=1$ state of the molecule D$_2$
 from Ref.~\cite{code} with the high precision theoretical results about the 
 electric field gradient at the nucleus of D$_2$, denoted by $q$ \cite{pavanello,jozwiak,komasa,puchalski}. 
 The fractional uncertainty of the most recent of these results reported in Ref.~\cite{puchalski}  -- 
 about $0.8\times10^{-4}$ -- comes from the fractional uncertainties of 
 the experimental value of $d$ and theoretical value of $q$,  $0.6\times10^{-4}$ and 
  $0.5\times10^{-4}$, respectively.
 Hyperfine spectroscopy of the D$_2^+$ ion offers the opportunity for an independent 
 spectroscopic determination of $Q_d$. On the example of the purely rotational $E2$
 transition $(vL)=(00)\to(v'L')=(02)$ we discuss the possibility to determine $Q_d$ using the 
 composite frequency method \cite{PRL2014,traceless,Alighanbari-2020}
 and the results of the present work as theoretical input,
 and estimate the accuracy of  $Q_d$ that can be achieved this way.
  
 We denote by ``composite frequency'' $\nu_c$ any linear combination of 
 resonance frequencies $\nu_i$ of
 transitions between the ro-vibrational states $(vL)$ and $(v'L')$ of D$_2^+$:
 \begin{equation}
 \nu_c=\sum\limits_{i=1}^N x_i \nu_i,
 \label{eq:compfreq}
 \end{equation}
 where $x_i$ are numerical coefficients, 
 normalized with $\sum\limits_{i=1}^Nx_i^2=1$, whose values are to be
 determined by imposing appropriate additional conditions. 
 The ``experimental'' value of the composite frequency is the linear combination of the 
 experimental data: $\nu_c^{ex}=\sum\limits_{i=1}^N x_i \nu_i^{ex}$,
 while the ``theoretical'' value $\nu_c^{th}=\sum\limits_{i=1}^N x_i\nu_i^{th}$
 is expressed in terms of the difference of the energy levels 
 of the initial (lower) and final (upper) states of the transitions as defined in 
 Eq.~(\ref{eq:hfsspl}): 
 \begin{eqnarray}
 &&\nu_i^{th}=(E^{(v'L')I'_iF'_iJ'_iJ'_{zi}}-E^{(vL)I_iF_iJ_iJ_{zi}})/h=
 \nu^{th}_{\rm spin-avg}+\Delta E^{(v'L')I'F'J'}-\Delta E^{(vL)IFJ},
  \label{eq:fc0}\\
 &&\nu^{th}_{\rm spin-avg}=(E^{{\rm (diag)}v'L'}-E^{{\rm (diag)}vL})/h.
 \nonumber
 \end{eqnarray} 
 The quantity $\nu_c^{th}$ is a known function of the 
 coefficients of the effective Hamiltonians 
 for the initial and final states $E_{in},E'_{in}, n=1,...,6$, and indirectly -- 
 of the physical constants involved, including $Q_d$:
 $\nu_c^{th}=\nu_c^{th}(E_{in},E'_{in},Q_d)$. 
 The value of $Q_d$ may be determined from the requirement that the theoretical value of the composite
 frequency be equal to the experimental one: 
 \begin{equation}
 \nu_c^{th}(E_{in},E'_{in},Q_d)=\nu_c^{ex}
 \label{eq:fc1}
 \end{equation}
 by resolving the latter with respect to $Q_d$:
 $Q_d=Q_d(E_{in},E'_{in},\nu_c^{ex})$.
 We introduce the notations $\Gamma_{in}=E_{n}(\partial \nu_i^{th}/\partial E_{n})$,
 $(x\cdot\Gamma_{n})=\sum\limits_{i=1}^N x_i \Gamma_{in}$ 
 (and similar for the ``primed'' symbols),
 where $\Gamma_{in}$ is the difference of the $\Gamma$'s defined in Eq.~(\ref{eq:uncDEhfs}) 
 for the final and initial state of the $i$-th transition:
 $\Gamma_{in}=\Gamma_n^{(v'L')I'F'J'}-\Gamma_n^{(vL)IFJ}$.  
 The absolute and fractional uncertainties $u(Q_d),u_r(Q_d)$ 
 of the quantity $Q_d$ arise from of the experimental uncertainty $u(\nu_c^{ex})$ of the  experimental composite frequency $\nu_c^{ex}$ and from the theory uncertainties of the Hamiltonian coefficients  $E_n$, $E'_n,n\le5$, $\bar E_6$, and $\bar E_6'$. We denote their fractional uncertainties by $u_n$,  $u'_n,  n\le6$, respectively. 
 Under the assumption that these uncertainties are 
 uncorrelated, we obtain the following estimate of the 
 fractional uncertainty $u_r(Q_d)$ of $Q_d$:
 \begin{eqnarray}
 &&u_r(Q_d)=\frac{u(Q_d)}{Q_d}=\sqrt{u_r^{th}(Q_d)^2+u_r^{ex}(Q_d)^2},\nonumber\\
 &&u_r^{th}(Q_d)=
 \frac
 {\sqrt{\sum\limits_{n=1}^6 
 \left((x\cdot\Gamma_{n})u_{n}\right)^2+
 \left((x\cdot\Gamma'_{n})u'_{n}\right)^2
 }}
 {\left|
 \left(x\cdot\Gamma_6\right)+
 \left(x\cdot\Gamma'_6\right)
 \right|},
 \ 
 u_r^{ex}(Q_d)=
 \frac
 {u(\nu_c^{ex})}
 {\left|
 \left(x\cdot\Gamma_6\right)+
 \left(x\cdot\Gamma'_6\right)
 \right|}
 \label{eq:urth+ex}
 \end{eqnarray}
 The coefficients $x_i$ are to be determined from the requirement that 
 the fractional uncertainty $u_r(Q_d)$ is minimal.
 Following the approach developed in \cite{Alighanbari-2020} we 
 restrict the search for minima by imposing the constraint 
 $\sum\limits_{i=1}^{N}x_i=0$ on the components of the
 vectors $x=\{x_i\}$. This suppresses the contribution 
 from higher-order QED and relativistic spin-independent effects to the 
 composite frequency. 
 
 As an illustration of the composite frequency approach, we estimate the accuracy 
 with which the value of $Q_d$ could be retrieved from a measurement 
 of the hyperfine structure of the $E2$ transition 
 $(vL)=(00)\to(v'L')=(02)$. (The numerical estimates for the vibrational transitions
 $(00)\to(v'2),v'\ge1$ are similar). 
 The spectral line has 11 strong (favored) hyperfine 
 components (see Table~\ref{tab:D2p_intesities}). For the initial state $(00)$ 
 all coefficients $E_{n}$ but $E_3$ vanish (see Table~\ref{tab:effs}). 
 The numerical values of $\Gamma'_{in},i=1,...,11,n=1,...,6$ and 
 the non-vanishing $\Gamma_{i3}$ are given in Table~\ref{tab:gamma}. 

 \begin{table}[h]
 \caption{Numerical values of the nonvanishing derivatives 
 $\Gamma_{i\alpha}\!=\!E_{\alpha}\left(\partial\nu_i/\partial E_{\alpha}\right)$ 
 and $\Gamma'_{i\alpha}\!=\!E'_{\alpha}\left(\partial\nu_i/\partial E'_{\alpha}\right)$
 for the strong (favored) hyperfine components of the $(00)\to(02)$ transition line, in MHz. 
 To ease comparison with Table~\ref{tab:even}, also given are the correspondent hyperfine shifts from 
 the spin-averaged E2-transition frequency $\Delta E^{\rm hfs}/h$, in MHz.
 $a[b]$ stands for $a\times10^b$.}
 \label{tab:gamma}
 \footnotesize
 \begin{tabular}{r@{\hspace{4mm}}r@{\hspace{4mm}}r@{\hspace{4mm}}r@{\hspace{4mm}}r@{\hspace{4mm}}
 r@{\hspace{4mm}}r@{\hspace{4mm}}r@{\hspace{4mm}}r}
 $i$ & \multicolumn{1}{c}{$\Delta E^{\rm hfs}$} & \multicolumn{1}{c}{$\Gamma'_{i1}$} 
 & \multicolumn{1}{c}{$\Gamma'_{i2}$} & \multicolumn{1}{c}{$\Gamma'_{i3}$} 
 & \multicolumn{1}{c}{$\Gamma'_{i4}$} & \multicolumn{1}{c}{$\Gamma'_{i5}$} & 
 \multicolumn{1}{c}{$\Gamma'_{i6}$} & \multicolumn{1}{c}{$\Gamma_{i3}$} \\ 
1& -40.018 &  -0.28721[+2] & 0.18094[$-$1] & 0.14130[+3]& -0.10064[+2] & 0.38755[$-$2]& -0.23231[$-$1]& -0.14253[+3]
\\
2& -32.093 & -0.32093[+2] & 0.00000[+0] &-0.27500[$-$5] & 0.50002[$-$7] & 0.25000[$-$6] & 0.29500[$-$5]& 0.00000[+0]
\\
3& -25.426 &  -0.20088[+2] & 0.14551[$-$1] & 0.13989[+3] &-0.27012[+1] & 0.15840[$-$2] &-0.94920[$-$2]&  0.14253[+3]
\\
4& -12.456 &  -0.15585[+2] &-0.11915[$-$1] &-0.21236[+3] & 0.17094[+1] & 0.10605[$-$2] &-0.63596[$-$2] & 0.21380[+3]
\\
5& -6.961 &  -0.85783[+1] & 0.82697[$-$2] & 0.13993[+3] & 0.42076[+1]& -0.12390[$-$2]  &0.74380[$-$2] & -0.14253[+3]
\\
6& -2.633 &  -0.21195[+1] & 0.13370[$-$2]& -0.21107[+3]& -0.32599[+1] &-0.24500[$-$2] & 0.14685[$-$1] &  0.21380[+3]
\\
7&   9.461 &  0.48872[+1] &-0.89350[$-$3]  &0.14122[+3] & 0.58717[+1]& -0.24026[$-$2] & 0.14401[$-$1] & -0.14253[+3]
\\
8&   11.084 &  0.93903[+1] & 0.11065[$-$1]& -0.21103[+3] &-0.10894[+1] &-0.24250[$-$3] & 0.14465[$-$2] &  0.21380[+3]
\\
9&  17.331 &   0.21395[+2] &-0.12809[$-$1]&  0.14228[+3] &-0.37906[+1] & 0.13415[$-$2]& -0.80425[$-$2] & -0.14253[+3]
\\
10&   21.395 &  0.21395[+2] & 0.00000[+0] & 0.49999[$-$5] & 0.10000[$-$6] &-0.60000[$-$6] &-0.68000[$-$5] & 0.00000[+0]
\\
11&   22.814 &  0.18024[+2] & 0.17128[$-$1]& -0.21244[+3] & 0.34302[+1] & 0.31675[$-$2] &-0.18990[$-$1] &  0.21380[+3]
  \end{tabular}
  \end{table}
 
 In absence of real spectroscopy data we assume in our considerations that the 
 experimental uncertainty of the composite frequency 
 $u(\nu_c^{ex})=\sqrt{\sum_{i=1}^N(x_i u(\nu_i^{ex}))^2}$ 
 (where $u(\nu_i^{ex})$ are the uncertainties of the individual hyperfine lines)
 satisfies $1<u(\nu_c^{ex})<170$ Hz and focus our interest to this range;
 in agreement with Eq.~(\ref{eq:uncE6}) we take $u_{n}=u'_{n}=\alpha^2\approx5\times10^{-5}$.
 In the range of interest, the uncertainty of $Q_d$ is determined -- and limited -- 
 by the experimental uncertainty: the contribution of the term $u_r^{ex}(Q_d)$ in Eq.~(\ref{eq:urth+ex})
 exceeds significantly $u_r^{th}(Q_d)$ (see Table~\ref{tab:xis}). 
 The dependence of  $u_r(Q_d)$ on $u(\nu_c^{ex})$ 
 is illustrated in Table~\ref{tab:uncertNN}. 
 The uncertainty $u_r(Q_d)$ decreases almost linearly with $u(\nu_c^{ex})$, 
 and monotonously with the growth of the number $N$ of hyperfine lines included in 
 the composite frequency $\nu_c$. However, beyond $N=6$ the advantage gained from adding 
 more components $\nu_i$ 
 is negligible (see Table~\ref{tab:uncertNN}). The optimal choice appears to be $N=6$;
 Table~\ref{tab:xis} lists the coefficients $x_i,i=1,...,6$ in the optimal linear combination
 $\nu_c=\sum_i x_i \nu_i,i\le6$ for a number of values of $u(\nu_c^{ex})$. 

 The above numerical estimates allow to outline the perspectives to determine $Q_d$ 
 by means of spectroscopy of D$_2^+$. 
 For an experimental uncertainty $u(\nu_c^{ex})=42.4$ Hz
 as in the recently reported high precision pure rotational spectroscopy of HD$^+$ 
 \cite{Alighanbari-2020}, the
 resulting fractional uncertainty of $Q_d$, $u_r(Q_d)\simeq0.0015$, is nearly an order of
 magnitude smaller that what was obtained in HD$^+$. 
 To distinguish the values of $Q_d$ reported in Refs.~\cite{komasa,jozwiak}, 
 on the one hand, and in Refs.~\cite{puchalski,pavanello}, on the other,
 the fractional uncertainty of $Q_d$ should be $u_r(Q_d)\lesssim10^{-3}$.
 Table~\ref{tab:uncertNN} shows that this could be achieved using the theoretical 
 results of the present work if $u(\nu_c^{ex})$ 
 does not exceed 30 Hz -- a level of experimental accuracy that has already been reached
 in HD$^+$ spectroscopy, albeit so far only on a single transition \cite{Alighanbari-2020}.   
 The accuracy of the extracted value of $Q_d$ would become competitive to the most precise 
 available results if $u(\nu_c^{ex})$ could be reduced to $\sim2$ Hz range. This is 
 close to the experimental precision needed to extract $Q_d$ by direct comparison 
 of the experimental and theoretical frequency of single hyperfine transitions 
 (see Sec.~\ref{sec:hfsnum}). However, only 
 the composite frequency approach 
 allows to reduce the theoretical error to a correspondingly low level.
 
 It should be stressed that the efficiency of this approach depends on the type of transition: 
 in transitions $(vL)\to(v'L')$ with $\Delta L=0$ the suppression of the theoretical uncertainty  
 $u_r^{th}(Q_d)$ is less pronounced. Also note that determining $Q_d$ from spectroscopy of 
 HD$^+$ with an uncertainty comparable to the values in Table~\ref{tab:uncertNN} would only be 
 possible with a significantly larger number $N$ of hyperfine components of the composite frequency
 compared to D$_2^+$ spectroscopy.
 
 \begin{table}[h]
 \caption{Achievable fractional uncertainty $u_r(Q_d)$ of the electric quadrupole moment of the deuteron
 $Q_d$ if retrieved from a composite frequency for the $E2$ transition $(00)\to(02)$  versus the number $N$ of hyperfine components involved, for experimental uncertainties $u(\nu_c^{ex})$
 between 170 and 1.3 Hz.
 $a[b]$ stands for $a\times10^b$.}
 \label{tab:uncertNN}
 \begin{tabular}{@{\hspace{4mm}}r@{\hspace{4mm}}r@{\hspace{4mm}}r@{\hspace{4mm}}r@{\hspace{4mm}}
 r@{\hspace{4mm}}r@{\hspace{4mm}}r@{\hspace{4mm}}r@{\hspace{4mm}}r@{\hspace{4mm}}}
 \hline\\
 & \multicolumn{8}{c}{Experimental uncertainty $u(\nu_c^{ex})$ (Hz)}\\
 \cline{2-9}
 \multicolumn{1}{c}{\vrule width 0pt height 11pt depth 2pt $N$} &
 \multicolumn{1}{l}{169.6} &
 \multicolumn{1}{l}{84.8} &
 \multicolumn{1}{l}{42.4} &
 \multicolumn{1}{l}{21.2} &
 \multicolumn{1}{l}{10.6} &
 \multicolumn{1}{l}{5.3} &
 \multicolumn{1}{l}{2.6} &
 \multicolumn{1}{l}{1.3} \\
\hline
 4 &   0.57[-2] &   0.34[-2] &   0.23[-2] &   0.19[-2] &   0.10[-2] &   0.60[-3] &   0.44[-3] &   0.39[-3] \\
 5 &   0.53[-2] &   0.28[-2] &   0.15[-2] &   0.85[-3] &   0.48[-3] &   0.29[-3] &   0.16[-3] &   0.10[-3] \\
 6 &   0.52[-2] &   0.27[-2] &   0.15[-2] &   0.80[-3] &   0.42[-3] &   0.22[-3] &   0.12[-3] &   0.76[-4] \\
 7 &   0.51[-2] &   0.27[-2] &   0.15[-2] &   0.79[-3] &   0.41[-3] &   0.21[-3] &   0.12[-3] & 0.75[-4]  
 \end{tabular}
 \end{table}

 \begin{table}[h]
 \caption{
 Theoretical, experimental, and overall fractional uncertainties  $u_r^{th}(Q_d)$, $u_r^{ex}(Q_d)$, 
 $u_r(Q_d)$ of the deuteron quadrupole moment (see Eq.~(\ref{eq:urth+ex})), and 
 coefficients $x_i$ of the optimal composite frequency $\nu_c$ of Eq.~(\ref{eq:compfreq})
 for $N=6$, as function of the experimental uncertainty $u(\nu_c^{ex})$.}
 \label{tab:xis}
 \footnotesize
\begin{tabular}{@{\hspace{2mm}}r@{\hspace{2mm}}r@{\hspace{2mm}}r@{\hspace{2mm}}r
@{\hspace{2mm}}l@{\hspace{2mm}}l@{\hspace{2mm}}l@{\hspace{2mm}}
 l@{\hspace{2mm}}l@{\hspace{2mm}}l@{\hspace{2mm}}}
 \hline\\
 \multicolumn{1}{c}{$u(\nu_c^{ex})$ } &
 \multicolumn{1}{c}{$u_r(Q_d)$} &
 \multicolumn{1}{c}{$u_r^{ex}(Q_d)$} &
 \multicolumn{1}{c}{$u_r^{th}(Q_d)$} &
 \multicolumn{6}{c}{Coefficients $x_i$ of the composite frequency $\nu_c$} 
 \\ 
 \vrule width 0pt height 11pt depth 2pt (Hz)\\\hline
 169.6 &  0.52[-2] & 0.51[-2] & 0.11[-2] & $x_{ 1}\!=\!\phantom{-}0.32087$ & $x_{ 4}\!=\!\phantom{-}0.14350$
  & $x_{ 5}\!=\!-0.10214$ & $x_{ 6}\!=\!-0.70800$ & $x_{ 7}\!=\!-0.21761$ & $x_{11}\!=\!\phantom{-}0.56338$ \\
 84.8 &  0.27[-2] & 0.26[-2] & 0.91[-3] & $x_{ 1}\!=\!-0.31117$ & $x_{ 4}\!=\!-0.18378$
  & $x_{ 5}\!=\!\phantom{-}0.15051$ & $x_{ 6}\!=\!\phantom{-}0.72653$ & $x_{ 7}\!=\!\phantom{-}0.15960$ & $x_{11}\!=\!-0.54169$ \\
 42.4 &  0.15[-2] & 0.14[-2] & 0.56[-3] & $x_{ 1}\!=\!\phantom{-}0.28355$ & $x_{ 2}\!=\!\phantom{-}0.11919$
  & $x_{ 4}\!=\!\phantom{-}0.12758$ & $x_{ 5}\!=\!-0.35445$ & $x_{ 6}\!=\!-0.69950$ & $x_{11}\!=\!\phantom{-}0.52363$ \\
 21.2 &  0.80[-3] & 0.77[-3] & 0.22[-3] & $x_{ 1}\!=\!-0.33803$ & $x_{ 2}\!=\!-0.25157$
  & $x_{ 3}\!=\!\phantom{-}0.21766$ & $x_{ 5}\!=\!\phantom{-}0.27110$ & $x_{ 6}\!=\!\phantom{-}0.64054$ & $x_{11}\!=\!-0.53970$ \\
 10.6 &  0.42[-3] & 0.41[-3] & 0.99[-4] & $x_{ 1}\!=\!-0.38008$ & $x_{ 2}\!=\!-0.24975$
  & $x_{ 3}\!=\!\phantom{-}0.32938$ & $x_{ 5}\!=\!\phantom{-}0.20048$ & $x_{ 6}\!=\!\phantom{-}0.61545$ & $x_{11}\!=\!-0.51547$ \\
  5.3 &  0.22[-3] & 0.21[-3] & 0.61[-4] & $x_{ 1}\!=\!-0.39541$ & $x_{ 2}\!=\!-0.24740$ &
   $x_{ 3}\!=\!\phantom{-}0.37397$ & $x_{ 5}\!=\!\phantom{-}0.16986$ & $x_{ 6}\!=\!\phantom{-}0.60123$ & $x_{11}\!=\!-0.50225$ \\
  2.7 &  0.12[-3] & 0.11[-3] & 0.54[-4] & $x_{ 1}\!=\!\phantom{-}0.11282$ & $x_{ 2}\!=\!\phantom{-}0.40157$
   & $x_{ 5}\!=\!-0.47272$ & $x_{ 6}\!=\!-0.61674$ & $x_{ 9}\!=\!\phantom{-}0.11895$ & $x_{11}\!=\!\phantom{-}0.45611$ \\
  1.3 &  0.76[-4] & 0.53[-3] & 0.54[-4] & $x_{ 1}\!=\!\phantom{-}0.11188$ & $x_{ 2}\!=\!\phantom{-}0.40259$
   & $x_{ 5}\!=\!-0.47320$ & $x_{ 6}\!=\!-0.61632$ & $x_{ 9}\!=\!\phantom{-}0.11976$ & $x_{11}\!=\!\phantom{-}0.45529$  
 \end{tabular}
 \end{table}
 
 \subsection{Determining the spin-averaged frequency by the composite frequency method}

 Similar to the above, one may also seek a composite frequency that contains the spin-averaged frequency but  minimizes the impact of the theory uncertainties of the HFS coefficients. This concept has been introduced in the  experimental work of Ref.~\cite{Alighanbari-2020} and in complementary form in Ref.~\cite{traceless}. 
 For D$_2^+$ this approach is particularly simple. As can be seen in Table~\ref{tab:gamma}, 
 lines 2 and 10 are practically only sensitive to the coefficient $E'_1$. Both the initial and final states 
 for these transitions are nearly pure, with strongly suppressed admixture of 
 other states, since the coupling through the terms with $E_2$, $E_3$, and $E_4$ of $H^{\rm eff}$ of  Eq.~(\ref{eq:heff}) vanishes for $I=0$, and $|E_{5,6}|\ll|E_1|$. This yields the approximate expressions
 for the hyperfine shift of energy levels of these states:
 $\Delta E^{(vL)0FJ}\approx E_1(J(J+1)-F(F+1)-L(L+1))/2$, which are accurate to better than $10^{-5}$.
 Therefore, one can determine the spin-averaged frequency by
 \begin{equation}
 \nu_{\rm spin-avg}^{ex}=
 \frac{2}{5}\nu_2^{ex}+
 \frac{3}{5}\nu_{10}^{ex}.
 \end{equation}
 The theoretical counterpart of this expression has 0.01 Hz uncertainty coming from 
 the spin structure theory, negligible compared to the uncertainty coming from the QED theory. 
 The same holds for the vibrational transitions $(0,0)\to(1,2)$ and $(0,0)\to(2,2)$.
 It is remarkable that 
 only two spin transitions suffice to yield the spin-averaged frequency; this may be a 
 significant advantage compared to HD$^+$.

 \section{\label{sec:Conclusion}Conclusion}

 The present paper reports a series of new theoretical
 results about the spectroscopy of the molecular ion D$_2^+$,
 including the most accurate to date calculations of the hyperfine structure
 of the lower excited ro-vibrational states with vibrational and
 rotational quantum numbers $v\le10$, $L\le4$.
 The correct theoretical treatment of the hyperfine
 structure is essential, considering that the experimental
 uncertainty of the measurement of the hyperfine structure has
 already reached the 20 Hz level in one of the molecular hydrogen
 ions \cite{Alighanbari-2020}.
 The paper also presents
 the first evaluation of the hyperfine structure of the $E2$
 ro-vibrational transitions, and a
 thorough consideration of the laser polarization effects in
 the laser spectroscopy of the resolved Zeeman components of
 the hyperfine transition lines.
 The work closely followed, but also further developed the
 formalism of Ref.~\cite{Korobov2018} for the study of the electric
 quadrupole transition spectrum of H$^+_2$.  
 The demonstrated efficiency of the composite frequency method underlines 
 the perspectives for precision spectroscopy on D$_2^+$ in the near future.
 Given the expected continued progress of experimental precision in trapped 
 molecular ion spectroscopy, this work provides strong motivation for 
 improving further the theoretical accuracy of the hyperfine coefficients 
 and of the QED energies. Apart from the well-known goals of deuteron-to-electron mass ratio 
 and electric quadrupole moment determination
 and test of QED, the spectroscopy of D$_2^+$
 specifically opens the possibility of searching for an
 anomalous force between deuterons \cite{salumbides2013}.

\begin{acknowledgments}
D.B. and P.D. gratefully acknowledge the support of Bulgarian
National Science Fund under Grant No. FNI 08-17. V.I.K.
acknowledges support from the Russian Foundation for Basic
Research under Grant No.~19-02-00058-a. S.S. acknowledges support
from the European Research Council (ERC) under the European Union's
Horizon 2020 research and innovation programme (grant agreement
No. 786306 ``PREMOL''). 
\end{acknowledgments}

\end{document}